\documentclass[english,preprint,5p]{elsarticle}
\usepackage[T1]{fontenc}
\usepackage[latin9]{inputenc}
\usepackage{color}
\usepackage{babel}
\usepackage{geometry}
\usepackage{amssymb}
\usepackage{graphicx}
\usepackage{array}
\usepackage{multirow}
\usepackage{float}
\usepackage[figuresright]{rotating}
\usepackage{multicol}
\usepackage{threeparttable}
\usepackage{epstopdf}

\makeatletter

\@ifundefined{showcaptionsetup}{}{%
 \PassOptionsToPackage{caption=false}{subfig}}
\usepackage{subfig}
\usepackage{lineno}

\makeatother

\begin{document}

\bibliographystyle{utcaps2.3_LdV}

\begin{frontmatter}

\title{Radiogenic and Muon-Induced Backgrounds in the LUX Dark Matter Detector}% Force line breaks with \\

\author[cwru]{D.S.~Akerib} 

\author[icl]{H.M.~Ara\'{u}jo} 

\author[sdsmt]{X.~Bai} 

\author[icl]{A.J.~Bailey} 

\author[umd]{J.~Balajthy} 

\author[yale]{E.~Bernard} 

\author[llnl]{A.~Bernstein} 

\author[cwru]{A.~Bradley} 

\author[usd]{D.~Byram} 

\author[yale]{S.B.~Cahn} 

\author[cwru,ucsb]{M.C.~Carmona-Benitez} 

\author[brwn]{C.~Chan} 

\author[brwn]{J.J.~Chapman} 

\author[usd]{A.A.~Chiller} 

\author[usd]{C.~Chiller} 

\author[cwru]{T.~Coffey} 

\author[icl]{A.~Currie} 

\author[lipc]{L.~de\,Viveiros} 

\author[umd]{A.~Dobi} 

\author[uedn]{J.~Dobson} 

\author[uofr]{E.~Druszkiewicz} 

\author[yale]{B.~Edwards}

\author[brwn,lbl]{C.H.~Faham} 

\author[brwn]{S.~Fiorucci} 

\author[ucd]{C.~Flores} 

\author[brwn]{R.J.~Gaitskell} 

\author[lbnl]{V.M.~Gehman} 

\author[ucl]{C.~Ghag} 

\author[cwru]{K.R.~Gibson} 

\author[lbnl]{M.G.D.~Gilchriese} 

\author[umd]{C.~Hall} 

\author[yale]{S.A.~Hertel} 

\author[yale]{M.~Horn} 

\author[brwn]{D.Q.~Huang} 

\author[ucb]{M.~Ihm} 

\author[ucb]{R.G.~Jacobsen} 

\author[llnl]{K.~Kazkaz} 

\author[umd]{R.~Knoche} 

\author[yale]{N.A.~Larsen} 

\author[cwru]{C.~Lee} 

\author[lipc]{A.~Lindote} 

\author[lipc]{M.I.~Lopes} 

\author[brwn]{D.C.~Malling\corref{cor1}} 

\author[tamu]{R.~Mannino} 

\author[yale]{D.N.~McKinsey} 

\author[usd]{D.-M.~Mei} 

\author[ucd]{J.~Mock} 

\author[uofr]{M.~Moongweluwan} 

\author[ucd]{J.~Morad} 

\author[uedn]{A.St.J.~Murphy} 

\author[ucsb]{C.~Nehrkorn} 

\author[ucsb]{H.~Nelson} 

\author[lipc]{F.~Neves} 

\author[ucd]{R.A.~Ott} 

\author[brwn]{M.~Pangilinan} 

\author[yale]{P.D.~Parker} 

\author[yale]{E.K.~Pease} 

\author[cwru]{K.~Pech} 

\author[cwru]{P.~Phelps} 

\author[ucl]{L.~Reichhart} 

\author[cwru]{T.~Shutt} 

\author[lipc]{C.~Silva} 

\author[lipc]{V.N.~Solovov} 

\author[llnl]{P.~Sorensen} 

\author[yale]{K.~O'Sullivan} 

\author[icl]{T.J.~Sumner} 

\author[ucd]{M.~Szydagis} 

\author[sdsta]{D.~Taylor} 

\author[yale]{B.~Tennyson} 

\author[sdsmt]{D.R.~Tiedt}

\author[ucd]{M.~Tripathi} 

\author[ucd]{S.~Uvarov} 

\author[brwn]{J.R.~Verbus} 

\author[ucd]{N.~Walsh} 

\author[tamu]{R.~Webb} 

\author[tamu]{J.T.~White}

\author[ucsb]{M.S.~Witherell} 

\author[uofr]{F.L.H.~Wolfs} 

\author[ucd]{M.~Woods} 

\author[usd]{C.~Zhang} 

\cortext[cor1]{Corresponding Author: David\_Malling@brown.edu}
\address[brwn]{Brown University, Dept. of Physics, 182 Hope St., Providence RI 02912, USA}
\address[cwru]{Case Western Reserve University, Dept. of Physics, 10900 Euclid Ave, Cleveland OH 44106, USA}
\address[hrvd]{Harvard University, Dept. of Physics, 17 Oxford St., Cambridge MA 02138, USA}
\address[icl]{Imperial College London, High Energy Physics, Blackett Laboratory, London SW7 2BZ, UK}
\address[lbl]{Lawrence Berkeley National Laboratory, 1 Cyclotron Rd., Berkeley CA 94720, USA}
\address[llnl]{Lawrence Livermore National Laboratory, 7000 East Ave., Livermore CA 94550, USA}
\address[lipc]{LIP-Coimbra, Department of Physics, University of Coimbra, Rua Larga, 3004-516 Coimbra, Portugal}
\address[sdsmt]{South Dakota School of Mines and Technology, 501 East St Joseph St., Rapid City SD 57701, USA}
\address[sdsta]{South Dakota Science and Technology Authority, Sanford Underground Research Facility, Lead, SD 57754, USA}
\address[tamu]{Texas A \& M University, Dept. of Physics, College Station TX 77843, USA}
\address[ucl]{University College London, Department of Physics and Astronomy, Gower Street, London WC1E 6BT, UK}
\address[ucb]{University of California Berkeley, Department of Physics, Berkeley CA 94720, USA}
\address[ucd]{University of California Davis, Dept. of Physics, One Shields Ave., Davis CA 95616, USA}
\address[ucsb]{University of California Santa Barbara, Dept. of Physics, Santa Barbara, CA, USA}
\address[uedn]{University of Edinburgh, SUPA, School of Physics and Astronomy, Edinburgh, EH9 3JZ, UK}
\address[umd]{University of Maryland, Dept. of Physics, College Park MD 20742, USA}
\address[uofr]{University of Rochester, Dept. of Physics and Astronomy, Rochester NY 14627, USA}
\address[usd]{University of South Dakota, Dept. of Physics, 414E Clark St., Vermillion SD 57069, USA}
\address[yale]{Yale University, Dept. of Physics, 217 Prospect St., New Haven CT 06511, USA}

\begin{abstract}

The Large Underground Xenon (LUX) dark matter experiment aims to detect rare low-energy interactions from Weakly Interacting Massive Particles (WIMPs). The radiogenic backgrounds in the LUX detector have been measured and compared with Monte Carlo simulation. Measurements of LUX high-energy data have provided direct constraints on all background sources contributing to the background model. The expected background rate from the background model for the 85.3 day WIMP search run is $(2.6\pm0.2_{\textrm{stat}}\pm0.4_{\textrm{sys}})\times10^{-3}$~events~keV$_{ee}^{-1}$~kg$^{-1}$~day$^{-1}$ in a 118~kg fiducial volume.  The observed background rate is $(3.6\pm0.4_{\textrm{stat}})\times10^{-3}$~events~keV$_{ee}^{-1}$~kg$^{-1}$~day$^{-1}$, consistent with model projections. The expectation for the radiogenic background in a subsequent one-year run is presented.

\end{abstract}

\begin{keyword}
LUX \sep dark matter \sep radioactive background \sep material screening \sep simulation
\end{keyword}

\end{frontmatter}

%\maketitle

%\linenumbers

%%%%%%%%%%%%%%%%%%%%%%%%%%%%%%%%%%%%%%%%%%%%%%%%%%%%%%%%%%%%%%%%%%%%%%%%%%%%%
\section{\label{sec:Introduction}Introduction}

The LUX experiment \cite{LUXNIM,LUXPRL} uses 370~kg of liquid Xe to search for nuclear recoil (NR) signatures from WIMP dark matter \cite{Blumenthal1984,Davis1985,Clowe2006}. The LUX detector reconstructs event energy, position, and recoil type through its collection of scintillation (S1) and electroluminescence (S2) signals. LUX seeks sensitivity to rare WIMP interactions at energies on the order of several keV. The extremely low WIMP interaction rate necessitates precise control of background event rates in the detector.

A particle that produces a WIMP search background in LUX must mimic a WIMP signature in several ways. WIMPs are expected to interact with Xe nuclei in the active region, creating a NR event. WIMP interactions will be single-scatter (SS) events, distributed homogeneously in the active region. The LUX WIMP search energy window is defined in the range 3.4--25~keV$_{nr}$, where the ``nr'' subscript denotes that the energy was deposited by a nuclear recoil \cite{LUXPRL}. This window captures 80\% of all WIMP interactions, assuming a WIMP mass of 100~GeV and standard galactic dark matter halo parameters as described in \cite{LUXPRL}.

The dominant background in the LUX WIMP search, which principally constrains the experimental sensitivity published for the 85.3~day run \cite{LUXPRL}, is low-energy electron recoil (ER) signatures in the Xe target. These events are generated through electromagnetic interactions from photons or electrons. The energy window for ER events differs from that of NR events due to differences in scintillation and ionization yield for each type of event. The 3.4--25~keV$_{nr}$ NR energy range has an S1 yield range equivalent to 0.9--5.3~keV$_{ee}$, where the ``ee'' subscript denotes an energy calibration for ER events. The ER energy range 0.9--5.3~keV$_{ee}$ is therefore taken as the WIMP search background range for ER events.

ER events are created mainly by $\gamma$~rays interacting in the 250~kg active volume. Gamma~rays are generated from the decay of radioisotope impurities in detector construction materials, with typical energies ranging from $\sim$100~keV to several~MeV. The dense liquid Xe target (2.9~g~cm$^{-3}$) attenuates $\gamma$~rays of these energies at the outer edge of the active region, with a mean free path on the order of several cm. Gamma~rays generated outside of the detector are suppressed below significance by the use of a 300~tonne water shield and 20~tonne external steel shield. The total water shielding thickness on all sides is $>$2.5~meters~water~equivalent (m.w.e.).

ER events are also generated by radioisotope decays within the Xe target itself. These isotopes are referred to as ``intrinsic.'' Intrinsic isotopes generate $\beta$~rays or X-rays that are fully absorbed within mm of the decay site. These isotopes are thoroughly mixed by convection and diffusion, and are distributed homogeneously in the active region. These energies of the $\beta$~rays or X-rays can fall within the 0.9--5.3~keV$_{ee}$ WIMP search energy range.

A subdominant background is expected from NR signatures from neutron scatters. Neutrons are generated internally in the detector through ($\alpha$,n) interactions in construction materials, and from spontaneous fission of $^{238}$U. These neutrons are generated with energies on the scale of MeV, with a mean free path of order 10~cm in liquid Xe. Neutrons are also generated from muon interactions in the laboratory and water shield. Muon-induced neutrons have energy at the GeV scale, with a mean free path in liquid Xe much longer than the size of the detector.

LUX uses S1 and S2 signal characteristics for multiple background rejection techniques. Scattering vertex positions in the detector are reconstructed with cm accuracy in XY, and mm accuracy in Z. This allows rejection of multiple scatter (MS) events, and enables the use of an inner fiducial region in which to conduct the WIMP search. The fiducial region excludes background events at the detector edges and maximizes WIMP signal-to-noise. Due to the limited $\gamma$~ray mean free path, together with the detector dimensions of 54~cm in height and 49~cm in diameter and use of an inner fiducial volume, the number of single-scatter $\gamma$~rays passing through the fiducial volume is four orders of magnitude less than the number of $\gamma$~rays with shallow penetration. The ratio of S2 to S1 also provides 99.6\%~discrimination against ER events on average over the WIMP search energy range.

This work details modeling and measurements of the LUX background rate from both electromagnetic and neutron sources. Monte Carlo simulation studies of all background components and direct measurement of signatures of these components in data are described in Sec.~\ref{sec:Background-Modeling}. The characterization of ER background rejection using the S2/S1 ratio is described in Sec.~\ref{sec:ER-NR-Disc}. Comparison of expected and measured low-energy background spectra is described in Sec.~\ref{sec:Comparison-with-lowE-data}.

%%%%%%%%%%%%%%%%%%%%%%%%%%%%%%%%%%%%%%%%%%%%%%%%%%%%%%%%%%%%%%%%%%%%%%%%%%%%%
\section{\label{sec:Background-Modeling}Background Modeling}

The LUX background model is comprised of multiple contributions. Each background source has been determined by direct measurements of LUX data, or from sampling measurements. These measurements are used to normalize Monte Carlo models of each background source. The Monte Carlo models are then used to project the expected low-energy background rate in the data. The details of the background model components, and their constraints from data, are described in this section.

%%%%%%%%%%%
\subsection{\label{sub:LUXSim}LUXSim}

\begin{figure}[htp]
\begin{centering}
\includegraphics[width=1\columnwidth]{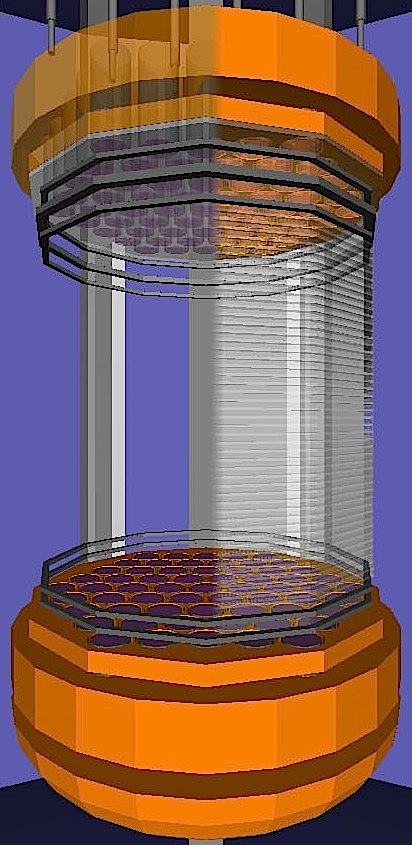}
\par\end{centering}
\caption{\label{fig:LUXSim-LUX-Geometry}Rendering of the LUX detector from the LUXSim simulation package. LUXSim is used for all studies of LUX radiogenic backgrounds. All geometry components are shown with semi-transparency. All high-mass detector construction materials are included in the simulation, as well as the external water shield. Details of all LUX geometry components are given in \cite{LUXNIM}. Plastic paneling and field shaping rings are cut away from the left side of the image to more clearly show the detailed structure of the PMT arrays. The cryostats are removed for clarity.}
\end{figure}

LUX background modeling work was performed using the Geant4-based LUXSim Monte Carlo simulation package \cite{Geant4,LUXSim}. The LUXSim package features a reconstruction of the full LUX detector based on CAD designs. The simulation geometry features all components with significant mass or relevance to photon collection modeling. The simulation also incorporates the LUX water shield, which thus enables accurate modeling of $\gamma$~ray and neutron scattering and moderation within the main detector. The high-fidelity representation plays an important role in the modeling of low-energy ER and NR background contributions from different detector components, as well as the effects of shielding in determining the $\gamma$~ray and neutron spectra incident at the active region. Simulation results were checked extensively using analytic calculations of signal distributions, based on particle trajectories and mean free paths in different detector components. A rendering of the LUXSim geometry is shown in Fig.~\ref{fig:LUXSim-LUX-Geometry}.

LUXSim also faithfully reproduces detector signals from ER and NR events. LUXSim incorporates the Noble Element Simulation Technique (NEST) package \cite{NEST2013}, along with several custom physics processes. These processes extend the Geant4 simulation to generate scintillation photons and ionization electrons in the active region, and are capable of reproducing LUX S1 and S2 waveforms. For all radiogenic backgrounds in this work, LUXSim was used to record only energy depositions, without the creation of scintillation and ionization signals. NEST is used with LUXSim to obtain the final low-energy S1 spectrum for direct comparison with measured data.

%%%%%%%%%%%%%%%%%%%%%%%%%%%%%%%%%%%%%%%%%%%%%%%%%%%%%%%%%%%%%%%%%%%%%%%%%%%%%
\subsection{Gamma~Rays from Construction Materials}

%%%%%%%%%%%
\subsubsection{Material Screening}

LUX construction materials were assayed for their radioactive content before use in detector construction. The materials were screened by high-purity Ge detectors at the Soudan Low-Background Counting Facility (SOLO) \cite{R11410} and Berkeley Oroville facility. These facilities in total screened $>$75~material samples in the course of the detector construction phase.

Screening results from counting at these facilities are summarized for high-mass ($\gtrsim$\,kg) construction materials in Tables~\ref{tab:Major-Counting-Description} and~\ref{tab:Major-Counting-Data}. The materials chosen represent >\,95\% of the detector dry mass contained within (and including) the cryostats. Photomultiplier tube (PMT) screening results are described in further detail in \cite{R11410}. Titanium cryostat material screening and cosmogenic activation studies are described in detail in \cite{LUXTi} and summarized in Sec.~\ref{sub:Cosmogenic-Activation}.

Screening results are reported for $^{238}$U and $^{232}$Th chain isotopes for all materials. For materials that commonly feature $^{40}$K or $^{60}$Co contamination, screening measurements or upper limits are also reported for these isotopes. No other radioisotope signatures were discovered during counting, with the exception of cosmogenic $^{46}$Sc in Ti.

% Screening material batch meta-info
\begin{sidewaystable*}[p]
\begin{centering}
\begin{tabular}{>{\centering}m{0.23\columnwidth}>{\centering}m{0.20\columnwidth}>{\centering}m{0.15\columnwidth}>{\centering}m{0.1\columnwidth}>{\centering}m{0.1\columnwidth}>{\centering}m{0.1\columnwidth}>{\centering}m{0.1\columnwidth}}
\hline 
\textbf{Component} & \textbf{Composition} & \textbf{Total Amount Used in LUX} & \textbf{Counting Facility} & \textbf{Counting Live Days} & \textbf{Quantity Counted} \tabularnewline
\hline 
PMTs &  & 122~PMTs & SOLO & See caption & 4 $\times$5 PMTs\tabularnewline
PMT bases &  & 122~bases & SOLO & 18 & 60~bases\tabularnewline
Field ring supports (inner panels) & HDPE & 18.0 kg & SOLO & 14 & 1.4~kg\tabularnewline
Field ring supports (outer panels) & HDPE & 15.5 kg & SOLO & 8.9 & 1.6~kg\tabularnewline
Reflector panels (main) & PTFE & 15.5 kg & SOLO & 9.5 & 2.4~kg\tabularnewline
Reflector panels (grid supports) & PTFE & 9.3 kg & SOLO & 6.7 & 2.0~kg\tabularnewline
Cryostats & Ti & 231 kg & Oroville & 13 & 8.0~kg\tabularnewline
Electric field grids & Stainless steel & 4.5 kg & SOLO & 7.3 & 37~kg\tabularnewline
Field shaping rings & Cu & 28 kg & SOLO & 9.0 & 15~kg\tabularnewline
PMT mounts & Cu & 169 kg & SOLO & 5.0 & 1.1~kg\tabularnewline
Weir & Cu & 3.2 kg & SOLO & 7.5 & 4.8~kg\tabularnewline
Filler chiller shield & Cu & 293 kg & SOLO & -- & -- \tabularnewline
Top shield & Cu & 121 kg & SOLO & -- & -- \tabularnewline
Superinsulation & Aluminized Mylar + polyester & 2.2 kg & SOLO & 13 & 0.6~kg\tabularnewline
Thermal insulation & HDPE & 6.0 kg & SOLO & 11 & 4.5~kg\tabularnewline
\hline 
\end{tabular}
\par\end{centering}
\caption{\label{tab:Major-Counting-Description}Description of high-mass LUX detector construction material samples assayed by $\gamma$~ray screening. The ``Total Amount Used in LUX'' column lists the summed number or mass of all units comprising the component. Counting results are listed in Table~\ref{tab:Major-Counting-Data}. PMT counting was performed in four batches with five PMTs each, with an average 9~live~days per batch \cite{R11410}. Components without a listed counting quantity were not counted, but are included because of their significant mass and proximity to the active region.}
\end{sidewaystable*}

% Screening material batch counting data
\begin{sidewaystable*}[p]
\begin{centering}
\begin{tabular}{cccccccc}
\hline 
\multirow{2}{*}{\textbf{Component}} & \multirow{2}{*}{\textbf{Counting Unit}} & \multicolumn{6}{c}{\textbf{Counting Results {[}mBq/unit{]}}}\tabularnewline
 &  & \textbf{$^{238}$U} & \textbf{$^{226}$Ra} & \textbf{$^{232}$Th} & \textbf{$^{40}$K} & \textbf{$^{60}$Co} & \textbf{Other}\tabularnewline
\hline 
PMTs & PMT & $<$22 & 9.5$\pm$0.6 & 2.7$\pm$0.3 & 66$\pm$6 & 2.6$\pm$0.2 & \tabularnewline
PMT bases & base & 1.0$\pm$0.4 & 1.4$\pm$0.2 & 0.13$\pm$0.01 & 1.2$\pm$0.4 & $<$0.03 & \tabularnewline
Field ring supports (inner panels) & kg &  & $<$0.5 & $<$0.35 &  &  & \tabularnewline
Field ring supports (outer panels) & kg &  & $<$6.3 & $<$3.1 &  &  & \tabularnewline
Reflector panels (main) & kg &  & $<$3 & $<$1 &  &  & \tabularnewline
Reflector panels (grid supports) & kg &  & $<$5 & $<$1.3 &  &  & \tabularnewline
Cryostats & kg & 4.9$\pm$1.2 & $<$0.37 & $<$0.8 & $<$1.6 & 4.4$\pm$0.3 ($^{46}$Sc) & \tabularnewline
Electric field grids & kg &  & $1.4\pm0.1$ & $0.23\pm0.07$ & $<$0.4 & $1.4\pm0.1$ & \tabularnewline
Field shaping rings & kg &  & $<$0.5 & $<$0.8 &  & $<$0.3 & \tabularnewline
PMT mounts & kg &  & $<$2.2 & $<$2.9 &  & $<$1.7 & \tabularnewline
Weir & kg &  & $<$0.4 & $<$0.2 &  & $<$0.17 & \tabularnewline
Superinsulation & kg & $<$270 & 73$\pm$4 & 14$\pm$3 & 640$\pm$60 &  & \tabularnewline
Thermal insulation & kg &  & 130$\pm$20 & 55$\pm$10 & $<$100 &  & \tabularnewline
\hline 
\end{tabular}
\par\end{centering}
\caption{\label{tab:Major-Counting-Data}Gamma radioassay data for high-mass LUX detector construction material samples. The $^{238}$U column lists measurements that are performed on $\gamma$~ray lines from isotopes in the $^{238}$U~early sub-chain, defined as all isotopes above $^{226}$Ra in the $^{238}$U decay chain. This measurement yields large errors and upper limits due to the very low branching ratios of these $\gamma$~rays, but provide the most direct measurement of $^{238}$U content. Measurements listed in the $^{226}$Ra column are performed on $\gamma$~ray lines from the $^{226}$Ra sub-chain, defined as isotopes including and below $^{226}$Ra in the $^{238}$U chain. The corresponding $\gamma$~rays have much higher intensities than those from the $^{238}$U~early sub-chain, and yield correspondingly lower activity limits. Radium-226 measurements are the typical results reported from LUX counting to determine $^{238}$U contamination, and were used for the $\gamma$~ray background initial projections in Fig.~\ref{fig:Gamma-Spectrum-Fit}. Both measurements are included for completeness (where available). Descriptions of all counting results are listed in Table~\ref{tab:Major-Counting-Description}. Reported errors are statistical only. Upper limits are at 90\% CL.}
\end{sidewaystable*}

%%%%%%%%%%%
\subsubsection{\label{sub:Cosmogenic-Activation}Cosmogenic Activation of Construction Materials}

The LUX detector was assembled and operated at the Sanford Surface Laboratory over a two~year period, before installation in the Davis Underground Laboratory. The operation of the LUX detector at the Sanford Surface Laboratory resulted in the cosmogenic activation of Ti and Cu detector construction materials. The activation products of concern with respect to detector backgrounds are $^{46}$Sc, generated in Ti, and $^{60}$Co, generated in Cu. Both of these isotopes have non-negligible half-lives ($^{46}$Sc 84~days; $^{60}$Co 5.3~years), and decay modes that can generate WIMP search ER backgrounds. Cosmogenic activation was stopped by moving the detector underground before the beginning of WIMP search running.

The LUX cryostats were selected from extremely low-radioactivity Ti stock. Counting results and activation studies are reported in \cite{LUXTi}. Titanium produces one transient radioisotope, $^{46}$Sc, from both muon capture and neutron spallation channels. Scandium-46 produces two simultaneous $\gamma$~rays with energies 889~keV and 1121~keV.

A 6.7~kg control Ti sample was used to estimate the total concentration of $^{46}$Sc produced in the LUX cryostats. The sample was screened at the SOLO facility after two years underground, and then transported by ground to the Sanford Surface Laboratory. The sample was activated over a six-month period before being transported by ground back to SOLO for re-analysis. Counting yielded a measurement of $4.4\pm0.3$~mBq~kg$^{-1}$ $^{46}$Sc. The measured decay rate was consistent with predictions based on the ACTIVIA simulation package \cite{ACTIVIA}, discussed in detail in Sec.~\ref{sub:Cosmogenic-Xenon-Production}.

Based on these measurements, the total $^{46}$Sc decay rate in the cryostats immediately after bringing the detector underground on July 12, 2012 was 1.3~Bq, conservatively assuming that the measured $^{46}$Sc content after the six-month exposure represented 75\% of the equilibrium activation value after the full LUX exposure. The elapsed time between moving LUX underground and beginning the 85.3~day WIMP search run was 284~days, and the 85.3~day WIMP search run was conducted over a period of 109~calendar days. The decay rate averaged over the 85.3~day run was $85\pm8$~mBq. The incident $\gamma$~ray flux from this source at the active region is below the level of measurement due to shielding from other internal materials, preventing a positive measurement of the $^{46}$Sc signature in data.

% Early studies of activated Ti determined that $\gamma$~ray backgrounds from $^{46}$Sc would become subdominant to $\gamma$~ray backgrounds from other detector materials before the start of the 85.3~day WIMP search run. For this reason, $^{46}$Sc is not included in the high-energy $\gamma$~ray background analysis.

The LUX detector uses 620~kg of ultra-low-activity oxygen-free high-thermal conductivity Cu in several construction components. These components include PMT array mounts, $\gamma$~ray radiation shields, thermal shields, and field shaping rings. A review of Cu activation studies in \cite{Cebrian2010} is used to estimate the $^{60}$Co production levels in LUX. A variation of a factor $\times$4 is found from all considered studies. From the distribution of results, the expected activation rate at sea level for $^{60}$Co in Cu is taken to be $62\pm29$~kg$^{-1}$~day$^{-1}$. The production rate is assumed to scale by a factor $\times$3.4 above that at sea level, tracking with the increase in muon-induced neutron flux at Sanford Surface Laboratory surface altitude \cite{Gordon2004}. The change in neutron spectrum at Sanford Surface Laboratory altitude relative to sea level is assumed to have a subdominant effect on the activation rate. The estimated production rate of $^{60}$Co in LUX internals is then $210\pm100$~kg$^{-1}$~day$^{-1}$. The estimated total exposure time for the Cu internals at the Sanford Surface Laboratory is 800~days, leading to a total decay rate of $1.0\pm0.5$~mBq~kg$^{-1}$ at the time the detector was taken underground. Counting results in Table~\ref{tab:Major-Counting-Data} for Cu components include upper limits on the presence of $^{60}$Co before the components arrived at Sanford.

%%%%%%%%%%%
\subsubsection{High-Energy Measurements and Model Fitting}

% No discussion of cuts is required here -- no cuts are used on LUXSim data

\begin{figure}[htp]
\begin{centering}
\includegraphics[width=1\columnwidth]{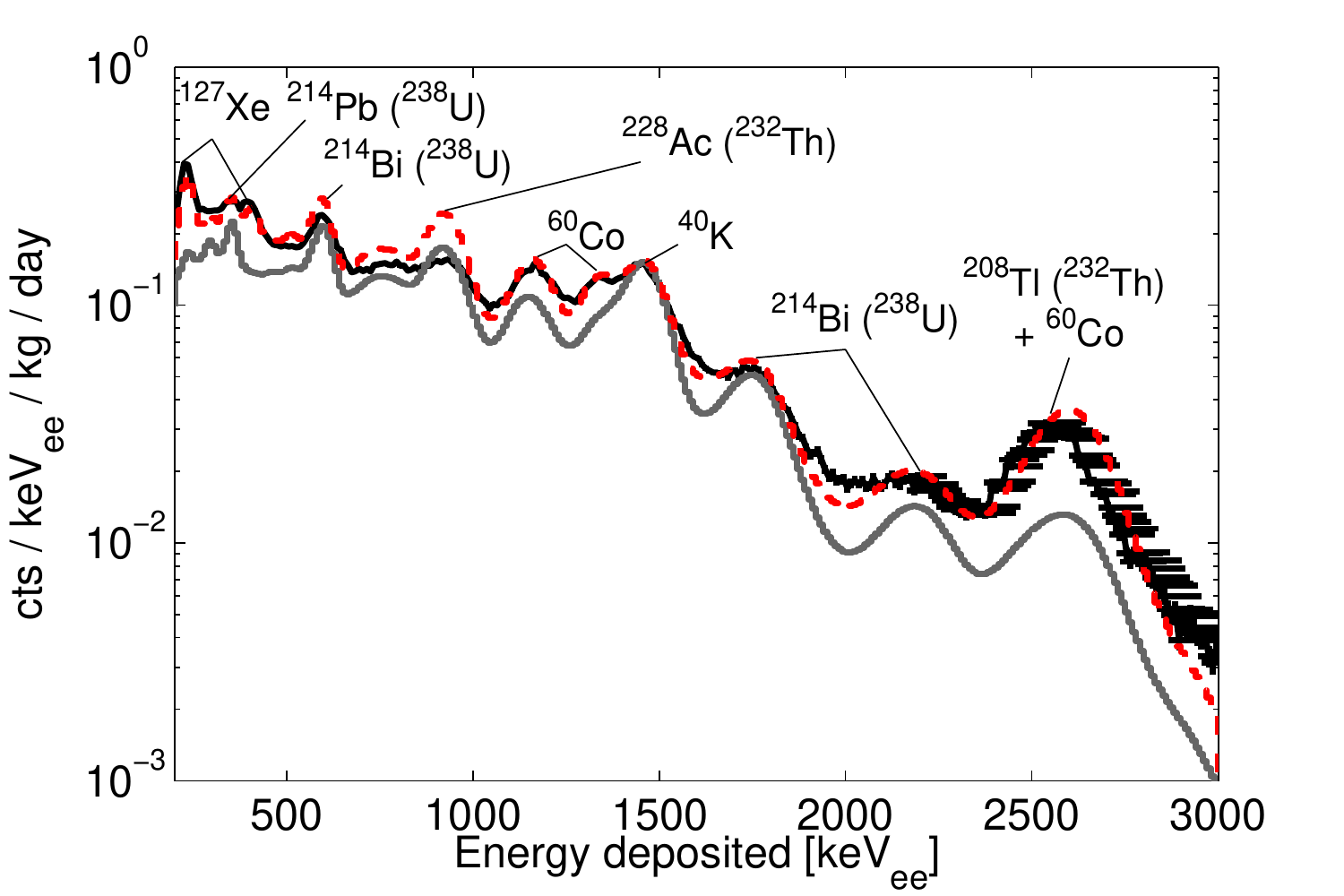}
\par\end{centering}
\caption{\label{fig:Gamma-Spectrum-Fit}Measured $\gamma$~ray spectrum in the LUX drift region (black), with peak identification labels. A 225~kg fiducial volume is used for the analysis, removing the top and bottom 2~cm of the drift region, and using no radial cut. Data includes both SS and MS events. Event energies are reconstructed from the combination of S1 and S2 signals. Horizontal error bars are shown, representing systematic uncertainties in energy reconstruction for high-energy events. Two simulation spectra are shown for comparison. A spectrum based on positive counting measurements alone is shown in gray solid. The spectrum with best-fit scaling for $^{238}$U, $^{232}$Th, $^{40}$K, and $^{60}$Co decays, with independent rates in top, bottom, and side regions of the detector, is shown as gray dashed (red, in color). Fitting was performed for energies above 500~keV$_{ee}$. Energies below 500~keV$_{ee}$ are shown to illustrate the continued agreement between $\gamma$~ray spectra and measured data below the fitting threshold. The spectrum shown has a lower bound at 200~keV$_{ee}$. Best-fit decay rates are listed in Table~\ref{tab:Best-fit-radioactivity-values}.}
\end{figure}

LUXSim $\gamma$~ray energy deposition spectra were generated as described in Sec.~\ref{sub:LUXSim}. The simulated spectra were compared with measured data to refine the estimate of $^{238}$U, $^{232}$Th, $^{40}$K, and $^{60}$Co concentration in detector internals. Energy spectra in the active region from each of these isotopes were simulated separately in the top, bottom, and lateral components adjacent to the active region. The decay rate of each isotope in each region was varied independently to obtain the best~fit to the measured $\gamma$~ray spectrum as a function of position in the active region. The fit included all energies above 500~keV$_{ee}$, avoiding influence from activated Xe spectra at peak energies of 408~keV$_{ee}$ and below. The measured $\gamma$~ray energy spectrum in the full drift region is compared with LUXSim model estimates in Fig.~\ref{fig:Gamma-Spectrum-Fit}. A two~cm volume is removed from the top and bottom to avoid contamination of the dataset from events in irregular field regions. Both the initial model based on screening data and the best~fit model are shown.

The best-fit simulation peak sizes show good agreement with observed data, with the exception of a 50\% predicted excess of the 969~keV line from $^{228}$Ac ($^{232}$Th chain). Agreement with the $^{228}$Ac peak cannot be found while preserving agreement with peaks at higher and lower energies, unless $^{232}$Th chain equilibrium is broken and the $^{228}$Ac rate is reduced independently. Equilibrium breakage for $^{228}$Ac alone is not a reasonable model, since $^{228}$Ac has a 6.2~hour half-life, and would regain its equilibrium concentration on this timescale. The deficit of $^{228}$Ac would suggest removal of the parent $^{228}$Ra from construction materials, e.g. during manufacturing processes. A detailed model of the isotope concentration was not investigated, since the presence of 50\% excess $^{228}$Ac in the model does not affect $\gamma$~ray or neutron background predictions in the WIMP search energy range.

The decay rates obtained from the best~fit model are given in Table~\ref{tab:Best-fit-radioactivity-values}. The total $^{238}$U, $^{232}$Th and $^{40}$K radioisotope content was found to be within one standard deviation of the predicted concentration from material screening. $^{60}$Co was found to have an excess consistent with a $1.7\pm1.0$~mBq concentration in Cu construction materials, in agreement with the predicted rate in Sec.~\ref{sub:Cosmogenic-Activation}.

\begin{figure}[htp]
\begin{centering}
\includegraphics[width=1\columnwidth]{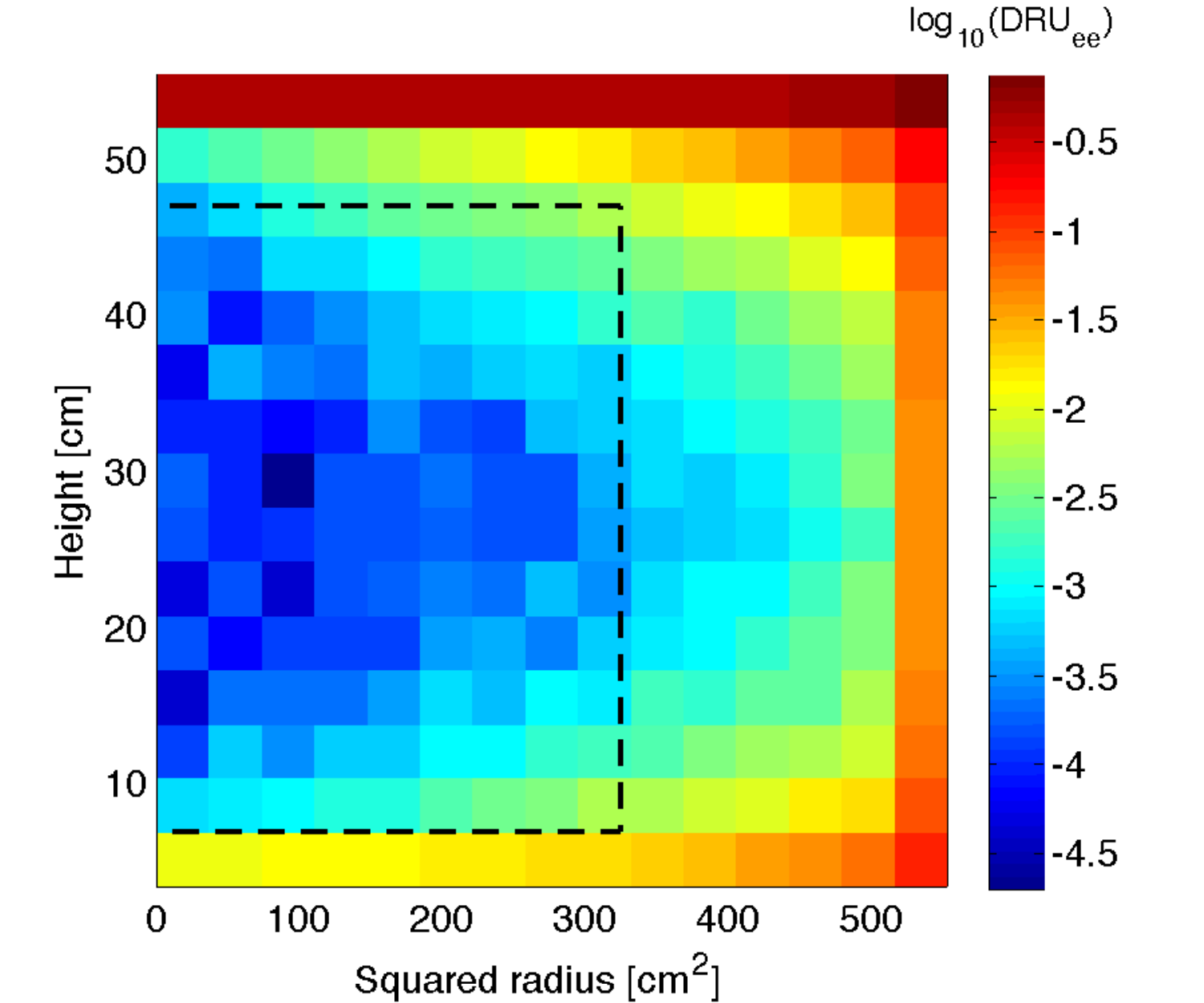}
\par\end{centering}
\caption{\label{fig:RZ-Gamma-BG}LUX $\gamma$~ray ER background density in the range 0.9--5.3~keV$_{ee}$ as a function of position, extrapolated from high-energy measurements based on Monte Carlo spectra. Rates are in units of $\log_{10}\left(\textrm{DRU}_{ee}\right)$. The 118~kg fiducial volume used in the 85.3~day WIMP search run is overlaid as the black dashed contour.}
\end{figure}

The ER background in the WIMP search energy range 0.9--5.3~keV$_{ee}$ is shown as a function of position from the $\gamma$~ray energy deposition spectra in Fig.~\ref{fig:RZ-Gamma-BG}. The decay rates are normalized to the best-fit results listed in Table~\ref{tab:Best-fit-radioactivity-values}. The $\gamma$~ray low-energy continuum is flat, and the background rates in units of DRU$_{ee}$ are independent of the exact energy window for the WIMP search.

\begin{table}
\begin{centering}
\begin{tabular}{ccccc}
\hline 
\multirow{2}{*}{\textbf{Region}} & \multirow{2}{*}{\textbf{Isotope}}	& \textbf{Screening} 		& \textbf{Best} \tabularnewline
						& 							& \textbf{Estimate {[}Bq{]}} & \textbf{Fit {[}Bq{]}} \tabularnewline
\hline 
\hline 
\multirow{4}{*}{Bottom} & $^{238}$U & $0.58\pm0.04$ & $0.62\pm0.16$ \tabularnewline
 & $^{232}$Th & $0.16\pm0.02$ & $0.23\pm0.06$ \tabularnewline
 & $^{40}$K & $4.0\pm0.4$ & $2.7\pm0.7$ \tabularnewline
 & $^{60}$Co & $0.16\pm0.01$ & $0.22\pm0.06$ \tabularnewline
\hline
\multirow{4}{*}{Top} & $^{238}$U & $0.58\pm0.04$ & $0.87\pm0.22$ \tabularnewline
 & $^{232}$Th & $0.16\pm0.02$ & $0.25\pm0.06$ \tabularnewline
 & $^{40}$K & $4.0\pm0.4$ & $3.8\pm1.0$ \tabularnewline
 & $^{60}$Co & $0.16\pm0.01$ & $0.30\pm0.08$ \tabularnewline
\hline
\multirow{4}{*}{Side} & $^{238}$U & $0.94\pm0.14$ & $0.22\pm0.06$ \tabularnewline
 & $^{232}$Th & $0.36\pm0.07$ & $1.5\pm0.38$ \tabularnewline
 & $^{40}$K & $1.4\pm0.1$ & $2.4\pm0.6$ \tabularnewline
 & $^{60}$Co & -- & $0.36\pm0.09$ \tabularnewline
\hline 
\end{tabular}
\par\end{centering}
\caption{\label{tab:Best-fit-radioactivity-values}Screening estimate and best-fit activity values for radioisotopes modeled in high-energy $\gamma$~ray analysis. Screening estimate values are taken from SOLO screening results for the PMTs and grids (corresponding to top and bottom regions), and superinsulation and plastic thermal insulation (side region). Materials with upper limits are not incorporated into the initial estimate. Errors on the best~fit values are estimated to be 25\%.}
\end{table}

%%%%%%%%%%%%%%%%%%%%%%%%%%%%%%%%%%%%%%%%%%%%%%%%%%%%%%%%%%%%%%%%%%%%%%%%%%%%%
\subsection{Cosmogenic Xenon Radioisotopes}

\subsubsection{\label{sub:Cosmogenic-Xenon-Production}Production Models}

The rate of production of noble element radioisotopes in Xe due to cosmic ray exposure was assessed using the ACTIVIA simulation package \cite{ACTIVIA}. The ACTIVIA code modeled isotope production in natural Xe after a 150~day exposure at sea level. Only noble elements were considered, as the LUX purification system is presumed to suppress the concentration of non-noble radioisotopes below significance \cite{Tritium,LUX01}.

The short-term exposure history of the LUX Xe is well known. From April 2012 to December 2012, the Xe was located at Case Western Reserve University (altitude 200~m) in a basement laboratory, where it was processed for Kr removal as discussed in Sec.~\ref{sub:Kr}. The Xe was shipped by ground to Sanford in separate batches and stored above ground (altitude 1.6~km), before being brought underground on January~30, 2013.  This adds up to roughly half the total Xe load in LUX having spent 49~days at Sanford Aboveground Laboratory altitude, and the other half having spent 7~days at that altitude. Reference~\cite{Gordon2004} provides some guidance for how to scale the muon-induced neutron flux and spectrum with altitude, which can be input into activation simulations. However, the effect of immediate surroundings in the lab can introduce an important systematic error in particular on the flux of thermal neutrons incident on the Xe. LUX does not have measurements of the thermal neutron flux at the various relevant locations. In the calculations below, the sea-level activation results from ACTIVIA were used as a starting point. Separate simulation results from 49~days and seven~days exposure at Lead, SD altitude for the appropriate Xe masses were added using the neutron flux correction factor of $\times$3.4 from~\cite{Gordon2004}. The uncertainty on the neutron flux and spectrum was then treated as a systematic. Because thermal neutron capture is the dominant process for the creation of several of the relevant isotopes, this results in a factor $\times$10 to $\times$100 uncertainty on the activities. ACTIVIA itself provides a factor $\times$2 uncertainty due to activation model parameters variations.

The final activity estimates are listed in Table~\ref{tab:Activated-isotope-rates}, for isotopes with concentrations $>$$10^{-5}$ after 90~days underground. All the predicted isotopes in the table are identified in initial low-background LUX data, and are discussed in Sec.~\ref{sub:Act-Xe-Meas}. A reasonable agreement for all isotopes (within a factor $\times$2) is found when applying a corrective factor of $\times$8 to all estimates, which may represent the variation in the thermal neutron flux. Those are the numbers reported in the table.

\begin{table}
\begin{centering}
\begin{tabular}{cccc}
\hline 
\multirow{2}{*}{{\bf Isotope}}	& {\bf Half-life}			& \multicolumn{2}{c}{{\bf Decay Rate {[}$\mu$Bq~kg$^{-1}${]}}} \tabularnewline
						& {\bf {[}Days{]}}		& {\bf Predicted}		& {\bf Observed} \tabularnewline
\hline 
\hline 
$^{127}$Xe				& 36					& 420				& $490\pm95$ \tabularnewline
$^{129m}$Xe				& 8.9					& 4.1					& $3.2\pm0.6$\tabularnewline
$^{131m}$Xe				& 12					& 25					& $22\pm5$ \tabularnewline
$^{133}$Xe				& 5.3					& 0.014				& $0.025\pm0.005$\tabularnewline
\hline 
\end{tabular}
\par\end{centering}
\caption{\label{tab:Activated-isotope-rates}Predicted Xe radioisotope activities in units of decay rate per kg detector target mass, produced from a combination of 150~day exposure at sea level and appropriate exposures at 1.6~km altitude (details in the text). Predicted and observed decay rates are listed after 90~days underground, where activation has ceased. Isotopes are shown which have a non-negligible concentration after 90~days underground. Activation rates are calculated using ACTIVIA. An overall factor $\times$8 is applied to all estimates in order to match the measurements, and is thought to represent the error on the thermal neutron flux. Errors on observations are based on uncertainty from peak fitting (Sec.~\ref{sub:Act-Xe-Meas}).}
\end{table}

\subsubsection{\label{sub:Act-Xe-Meas}Measurement in LUX Data}

\begin{figure}[htp]
\begin{centering}
\includegraphics[width=1\columnwidth]{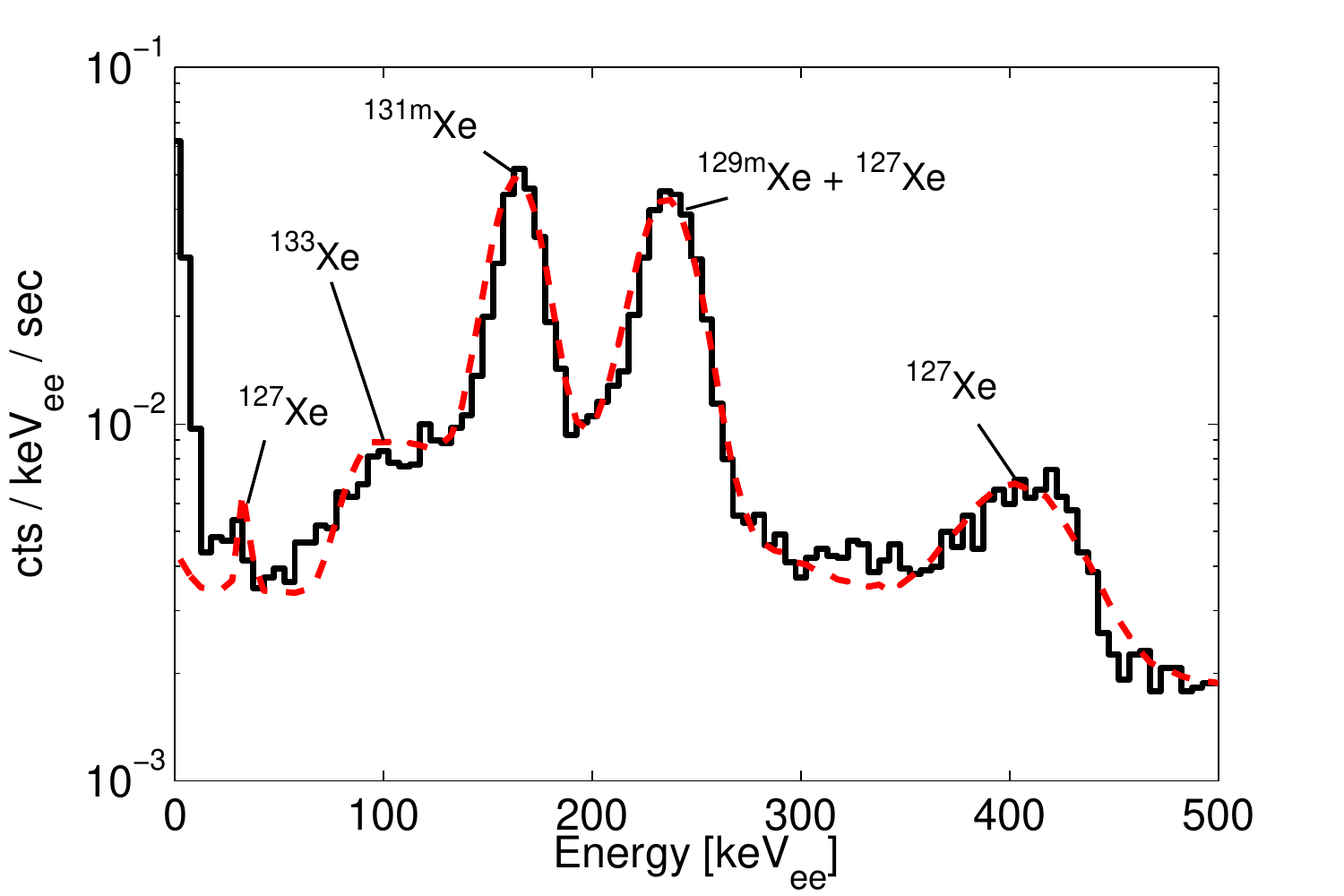}
\par\end{centering}
\caption{\label{fig:Activated-Xenon}Early zero-field LUX data taken 12~days after bringing the Xe underground, featuring peaks from cosmogenically activated Xe isotopes. The measured spectrum is shown as the black histogram. The best-fit spectrum for cosmogenic Xe isotopes plus exponential background is shown as the dashed curve (gray in print, red online). The simulation spectrum is comprised of $^{127}$Xe, $^{129m}$Xe, $^{131m}$Xe, and $^{133}$Xe components, with decay rates of 2.7, 3.6, 4.4, and 3.6~mBq~kg$^{-1}$ respectively. Simulated peak resolution is measured from the $^{129m}$Xe and $^{131m}$Xe peaks and extrapolated as $\sqrt{E}$.}
\end{figure}

Signatures of $^{129m}$Xe and $^{131m}$Xe isotopes were originally identified in LUX surface run data \cite{LUXSurfaceRun}. After the start of underground operations, the $^{129m}$Xe and $^{131m}$Xe isotope concentration estimates were refined. Two additional activation isotopes, $^{127}$Xe and $^{133}$Xe, were also identified in pre-WIMP run LUX data. The energy spectrum of these isotopes, taken from zero-field data 13~days after the Xe was moved underground and 70~days before the start of the WIMP search run, is shown in Fig.~\ref{fig:Activated-Xenon}. A fitted simulation spectrum is overlaid with contributions from the four Xe cosmogenic isotopes plus an exponential contribution from $\gamma$~ray Compton background. The best-fit decay rates for these spectra correspond to $2.7\pm0.5$~mBq~kg$^{-1}$ ($^{127}$Xe), $3.6\pm0.7$~mBq~kg$^{-1}$ ($^{129m}$Xe), $4.4\pm0.9$~mBq~kg$^{-1}$ ($^{131m}$Xe), and $3.6\pm0.7$~mBq~kg$^{-1}$ ($^{133}$Xe) after 13~days underground.

\begin{figure}[htp]
\begin{centering}
\subfloat[]{\includegraphics[width=1\columnwidth]{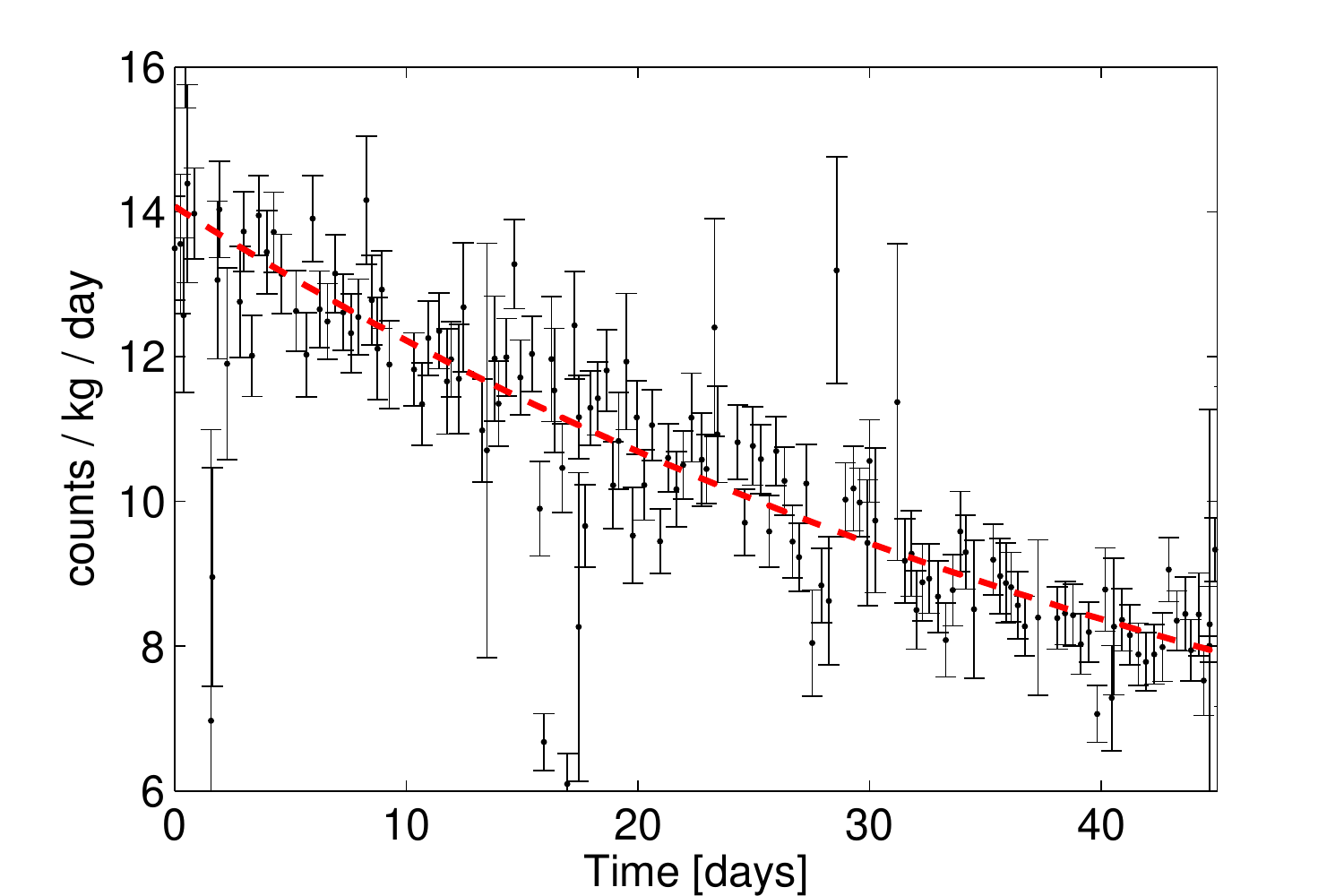}
}

\subfloat[]{\includegraphics[width=1\columnwidth]{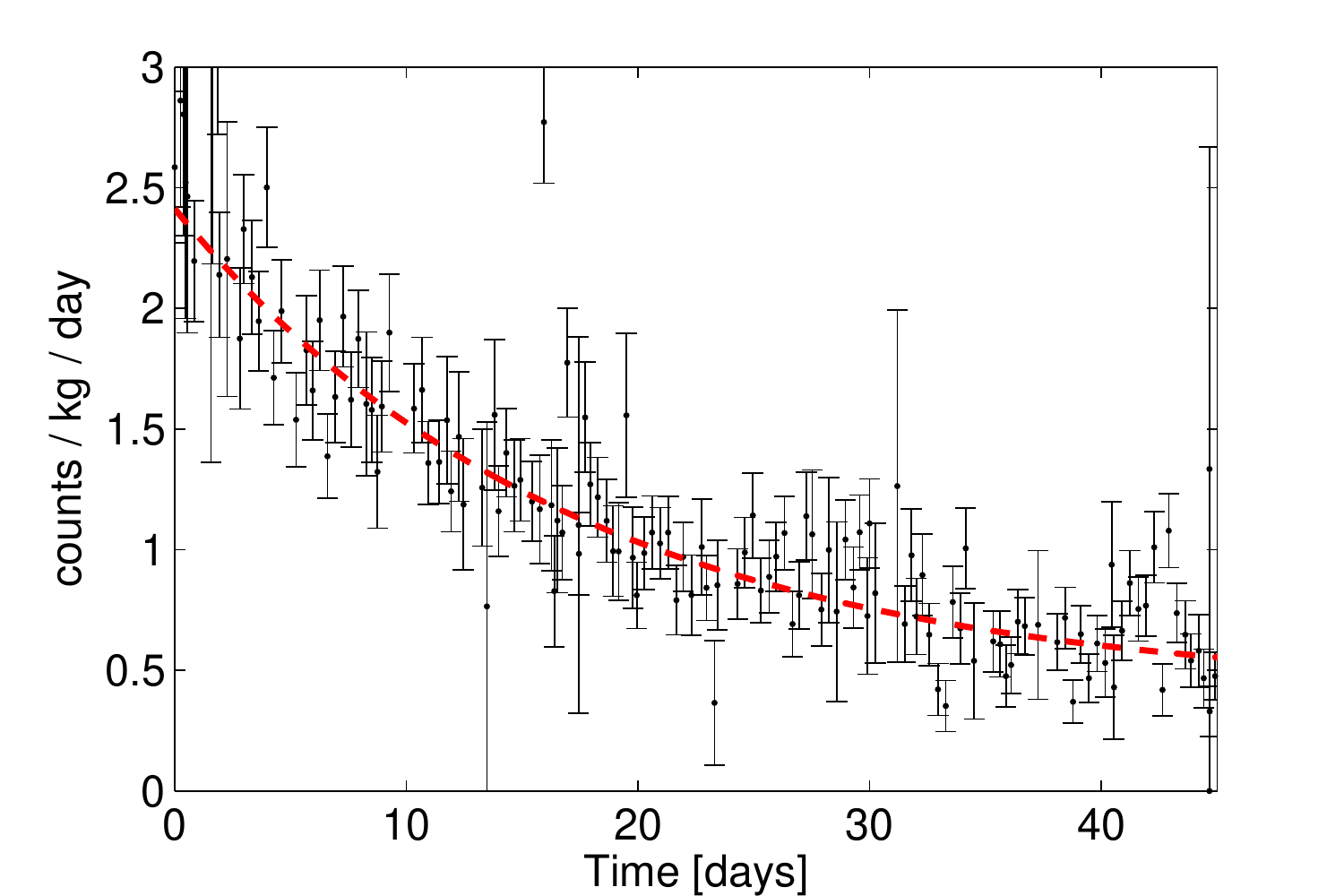}
}
\par\end{centering}
\caption{\label{fig:Activated-Xenon-Decay-vs-Time}Activated Xe peak rates as a function of time, measured over part of the 85.3~day WIMP search run, beginning May~1, 2013. Peaks are measured for (a) $^{127}$Xe, at 375~and 410~keV$_{ee}$, and (b) $^{131m}$Xe, at 164~keV$_{ee}$. Best-fit exponential functions are shown with dashed lines (red, in color). The exponential slope of each function is fixed to the corresponding literature half-life listed in Table~\ref{tab:Activated-isotope-rates}, and is not varied with the fit. Error bars are statistical.}
\end{figure}

The decay rates of $^{127}$Xe and $^{131m}$Xe were measured by the decay of the 375~keV$_{ee}$ and 164~keV$_{ee}$ peaks, respectively, over the course of the 118~day WIMP search run. The measured peak rates and best-fit models are shown in Fig.~\ref{fig:Activated-Xenon-Decay-vs-Time}. The decay fits yield concentrations of $115\pm20$~$\mu$Bq~kg$^{-1}$~$^{127}$Xe and $7.3\pm1.5$~$\mu$Bq~kg$^{-1}$~$^{131m}$Xe, averaged over the second half of the 85.3~day WIMP search run. The concentrations calculated from the peak decays agree with initial projections from early zero-field data estimates.

The measured decay rates of all identified Xe radioisotopes are listed in Table~\ref{tab:Activated-isotope-rates}. The measured decay rates are within a factor $\times$2 of predictions, if one also applies a corrective factor $\times$8 to all predicted values. This is thought to be related to the uncertainty on the thermal neutron flux, as well as uncertainties inherent to the simulation model. Further studies and measurement are planned in order to try and control these uncertainties.

Useful energy calibration points were found from $^{127}$Xe, $^{131m}$Xe and $^{129m}$Xe. The 5.2~day half-life of $^{133}$Xe rendered this isotope unmeasurable by the start of WIMP search running. The only activated Xe isotope capable of generating a significant WIMP search background is $^{127}$Xe.

\subsubsection{\label{sub:Xe127}$^{127}$Xe Backgrounds}

Decays of $^{127}$Xe generate a low-energy ER background in the LUX 85.3~day WIMP search data. Xenon-127 decays via electron capture, resulting in an orbital vacancy. The vacancy is filled by electron transitions from higher orbitals, resulting in an X-ray or Auger electron cascade. In the case of $^{127}$Xe, the capture electron comes from the K~shell with 85\% probability \cite{ENSDF}, resulting in a cascade with total energy 33~keV. A further 12\% of captures come from the L~shell, with a cascade of total energy 5.2~keV. The remainder of the decays come from the higher shells, with energy $\leq$1.2~keV. The L~shell decay energy is at the upper edge of the WIMP search window, with $\sim$50\% of all decays falling inside the window. The lower-energy decays occur at the low end of the WIMP search range, and are conservatively estimated to fall inside the search window with 100\% efficiency.

The daughter $^{127}$I nucleus is left in the 203~or 375~keV excited states, with 53\%~or 47\%~respective probability. There is a 17\% probability of decay from the 375~keV state to ground by $\gamma$~ray emission. This $\gamma$~ray has a mean free path of 2.6~cm in liquid Xe, and can potentially escape from the active region.

A low-energy EM signature arises when the low-energy X-ray deposition occurs in coincidence with the escape of the 375~keV $\gamma$~ray. The $^{127}$Xe background rate is dependent on the escape probability of the $\gamma$~ray. Therefore, although the $^{127}$Xe concentration in the active region is homogeneous, the background from $^{127}$Xe is seen to fall exponentially with distance from the active region edges. The estimated $^{127}$Xe background in the 118~kg fiducial volume, estimated from LUXSim studies, is $0.5\pm0.02_{\textrm{stat}}\pm0.1_{\textrm{sys}}$~mDRU$_{ee}$, averaged over the 85.3~day WIMP search run. This background decreases with a 36~day half-life, and will not be present for the one-year run beginning in 2014.

%%%%%%%%%%%%%%%%%%%%%%%%%%%%%%%%%%%%%%%%%%%%%%%%%%%%%%%%%%%%%%%%%%%%%%%%%%%%%
\subsection{Radon}

\subsubsection{Identification of Radon Daughters in LUX Data}

The decay of $^{222}$Rn and $^{220}$Rn daughters generates a low-energy ER background in LUX data. Radon isotopes decay through several short-lived daughter stages. $^{222}$Rn generates $^{214}$Pb and $^{214}$Bi. $^{220}$Rn generates $^{212}$Pb, which decays with a scheme similar to $^{214}$Pb. These isotopes undergo ``naked'' or ``semi-naked'' $\beta$~decay. Naked $\beta$~decay refers to $\beta$~ray emission without accompanying EM emission that could veto the event. Semi-naked $\beta$~decay refers to $\beta$~ray emission accompanied by emission of a high-energy $\gamma$~ray, which can potentially escape the active region. If the $\gamma$~ray escapes the active region, the $\beta$~ray is not tagged as a background event.

Radon daughters were identified in LUX data through $\alpha$~decay signatures. Alpha particles are clearly distinguished in LUX data by their large S1 signal sizes, ranging from $4\times10^4$~--~$9\times10^4$~photoelectrons (phe). The S1 pulse sizes from $\alpha$~particle events are much larger than S1 pulses from $\gamma$~ray events, which reach up to $1.5\times10^4$~phe. Alpha particles produce a clear signature in LUX data, which can be used to characterize the $^{222}$Rn and $^{220}$Rn chain decay rates and distributions in the active region. These isotopes are the only sources of $\alpha$~decays in LUX.

\begin{figure}[htp]
\begin{centering}
\includegraphics[width=1\columnwidth]{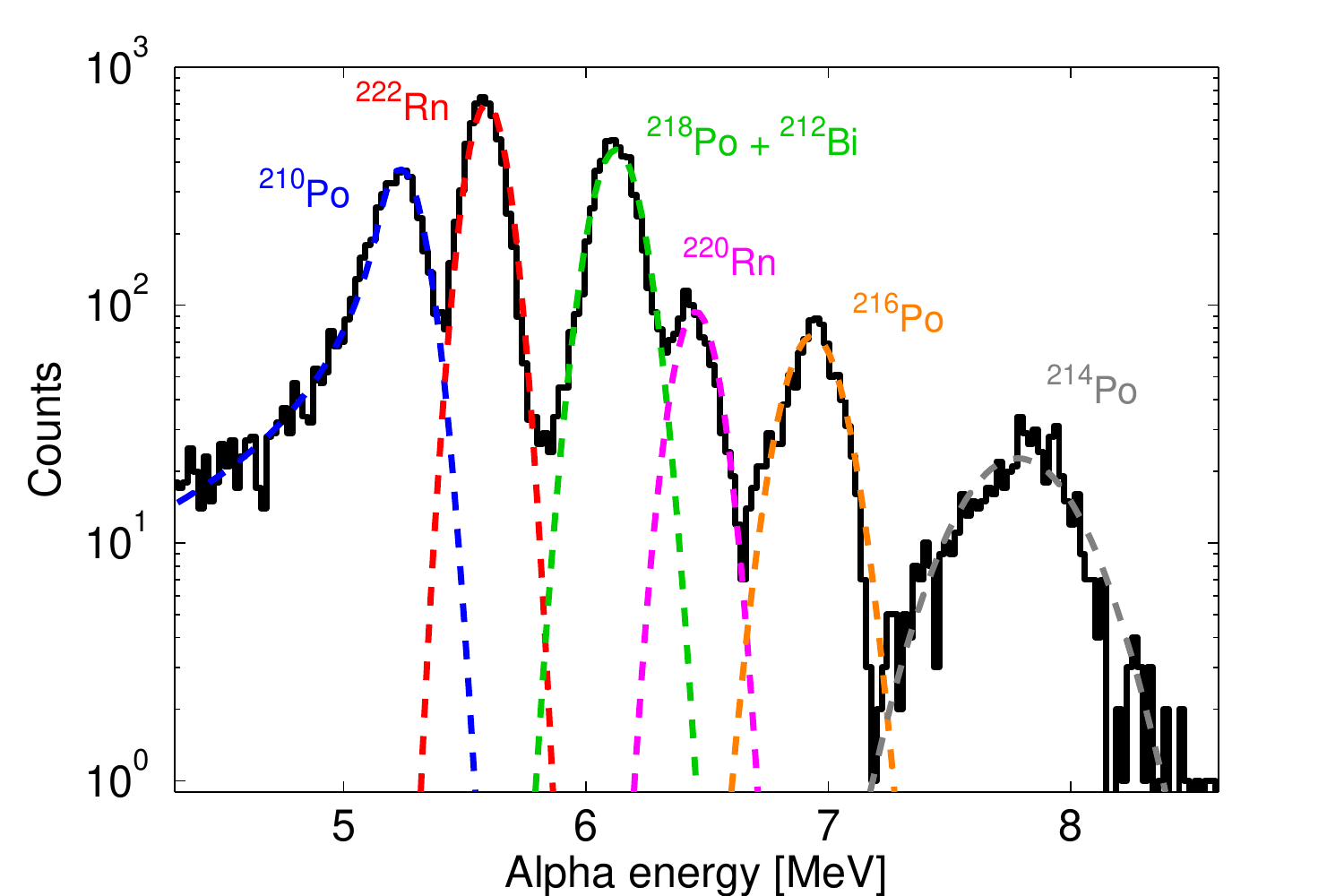}
\par\end{centering}
\caption{\label{fig:Alpha-Peaks-S1}Signatures of $\alpha$~particle decays in LUX WIMP search data in the active region. Energy is reconstructed from S1 measurements, calibrated on the observed location of the $^{222}$Rn peak. Counts are collected over 6~livedays, spaced periodically over the course of the 85.3~day WIMP search run. Measured data (black) are fitted with Gaussian curves, with the exception of the $^{210}$Po peak which is fit with a Crystal Ball distribution (see text for discussion). Fitted curves are shown in gray dashed (peaks shown in color online). Fitted curve peak values are fixed at the Q-values of the $\alpha$~decays for all isotopes, with the exception of $^{210}$Po where the peak value is the mean $\alpha$~particle energy.}
\end{figure}

Identified $\alpha$~particle peaks are shown in Fig.~\ref{fig:Alpha-Peaks-S1}. The six $\alpha$~particle peaks are fit to a sum of five Gaussians and one Crystal Ball distribution \cite{Gaiser:1982}, which is a Gaussian with a power law tail and is characteristic of lossy processes. The Crystal Ball distribution is used for the $^{210}$Po peak, to characterize partial $\alpha$~particle energy loss from transit through materials in contact with the active region. The peak means are scaled for a best~fit to literature values of $\alpha$~particle energies from $^{222}$Rn and $^{220}$Rn daughters. The total measured daughter rates were taken from the best-fit peak areas. The radon daughter isotopes, $\alpha$~particle energies, half-lives, and measured decay rates are summarized in Table~\ref{tab:Alpha-Rates}.

The reconstruction efficiency for short-lived isotopes is limited due to overlap with the parent event. The efficiency for $^{214}$Po event reconstruction was estimated at 52\%, based on the efficiency of S2 pulse separation between the $^{214}$Po and the parent $^{214}$Bi $\alpha$~particle events. Poor reconstruction efficiency was found for overlapping $^{212}$Bi~/~$^{212}$Po events, which comprise virtually all $^{212}$Po decay events due to its 0.3~$\mu$s half-life. No $^{212}$Po measurement is reported for this reason.

The reconstructed energies of all $\alpha$~particle decays are taken to be the total Q-values of the decays, accounting for both the $\alpha$~particle and the recoiling daughter nucleus. The exception is $^{210}$Po, which decays on detector surfaces. For $^{210}$Po, it is assumed that the reconstructed energy is the $\alpha$~particle energy only. If the $^{210}$Po $\alpha$~particle is detected in the active region, then the recoiling $^{206}$Pb nucleus becomes further embedded in the material surface and does not produce a visible signal. If the $^{206}$Pb nucleus recoils into the active region, then the $\alpha$~particle will be emitted into the PTFE and will not contribute to the observed 5.3~MeV peak.

% Table of isotope, energies, half-life, measured event rate, and general position
\begin{table*}
 	\begin{center}
 	\begin{threeparttable}
	\begin{tabular}{cccccc}
	\hline
	\textbf{Decay} & \multirow{2}{*}{\textbf{Isotope}} & \textbf{Energy} & \textbf{Measured} & \multirow{2}{*}{\textbf{Half-life}} & \multirow{2}{*}{\textbf{Event Rate {[}mHz{]}}} \\ 
	\textbf{Chain} & & \textbf{{[}MeV{]}} & \textbf{Energy {[}MeV{]}} & & \\
	\hline
	\hline
	\multirow{5}{*}{$^{238}$U} & $^{222}$Rn & 5.59 & $5.59\pm0.08$ & 3.8~d & $17.9\pm0.2$ \\
	 & $^{218}$Po & 6.16 & $6.12\pm0.10$ & 3.1~min & $14.4\pm0.2$ \\
	 & $^{214}$Po & 7.84 & $7.80\pm0.2$ & 160~$\mu$s & $3.5\pm0.1$ \\
	& \multirow{2}{*}{$^{210}$Po} & \multirow{2}{*}{5.30} & \multirow{2}{*}{$5.22\pm0.09$} & \multirow{2}{*}{140~d} & $14.3\pm0.2$ (on walls) \\
	 & & & & & $7.2\pm0.2$ (on cathode) \\
	\hline
	\multirow{4}{*}{$^{232}$Th} & $^{220}$Rn & 6.41 & $6.47\pm0.09$ & 56~s & $2.6\pm0.1$ \\
	 & $^{216}$Po & 6.91 & $6.95\pm0.1$ & 0.15~s & $2.8\pm0.1$ \\
	 & $^{212}$Bi & 6.21& $6.12\pm0.10$ & 61~min & $14.4\pm0.2$ \\
	 & $^{212}$Po & 8.83 & -- & 0.30~$\mu$s & -- \\
	\hline
	\end{tabular}
	\end{threeparttable}
	\end{center}
	\caption[Alphas found in the LUX detector, their measured energies, parent decay chain, and measured activities during the 85.3~day WIMP search run.]{\label{tab:Alpha-Rates}Radon chain daughter isotopes measured in the LUX active volume during the 85.3~day WIMP search run. Measurements are collected over 6~livedays, spaced periodically over the course of the 85.3~day WIMP search run. The known energies are the decay Q-values, except in the case of $^{210}$Po, where the listed known energy is the $\alpha$~particle emission energy. $^{218}$Po and $^{212}$Bi are too close in energy to be resolved as separate peaks. Errors are statistical, and are reported at $\pm34\%$.}
\end{table*}

\subsubsection{Radon Daughter Backgrounds}

Radon daughters that generate low-energy ER backgrounds are not directly countable in LUX data. However, their decay rates can be bounded based on measurements of parent and daughter $\alpha$~decays. The decay rates of $^{214}$Bi and $^{214}$Pb are bounded by the measured $^{218}$Po and $^{214}$Po rates, yielding a range of 3.5--14~mBq in the LUX active region. The $^{212}$Pb rate has an upper bound of $<$2.8~mBq from $^{216}$Po, and is assumed to be further removed from the active region due to its 11~hour half-life. $^{214}$Bi is also removed from consideration as an ER background source due to the 160~$\mu$s half-life of the daughter $^{214}$Po, which creates a 90\% probability of overlap with the $^{214}$Po $\alpha$~decay within the LUX 1~ms event window. The primary isotope of concern with respect to ER backgrounds is therefore $^{214}$Pb.

\begin{figure}[htp]
\begin{centering}
\includegraphics[width=1\columnwidth]{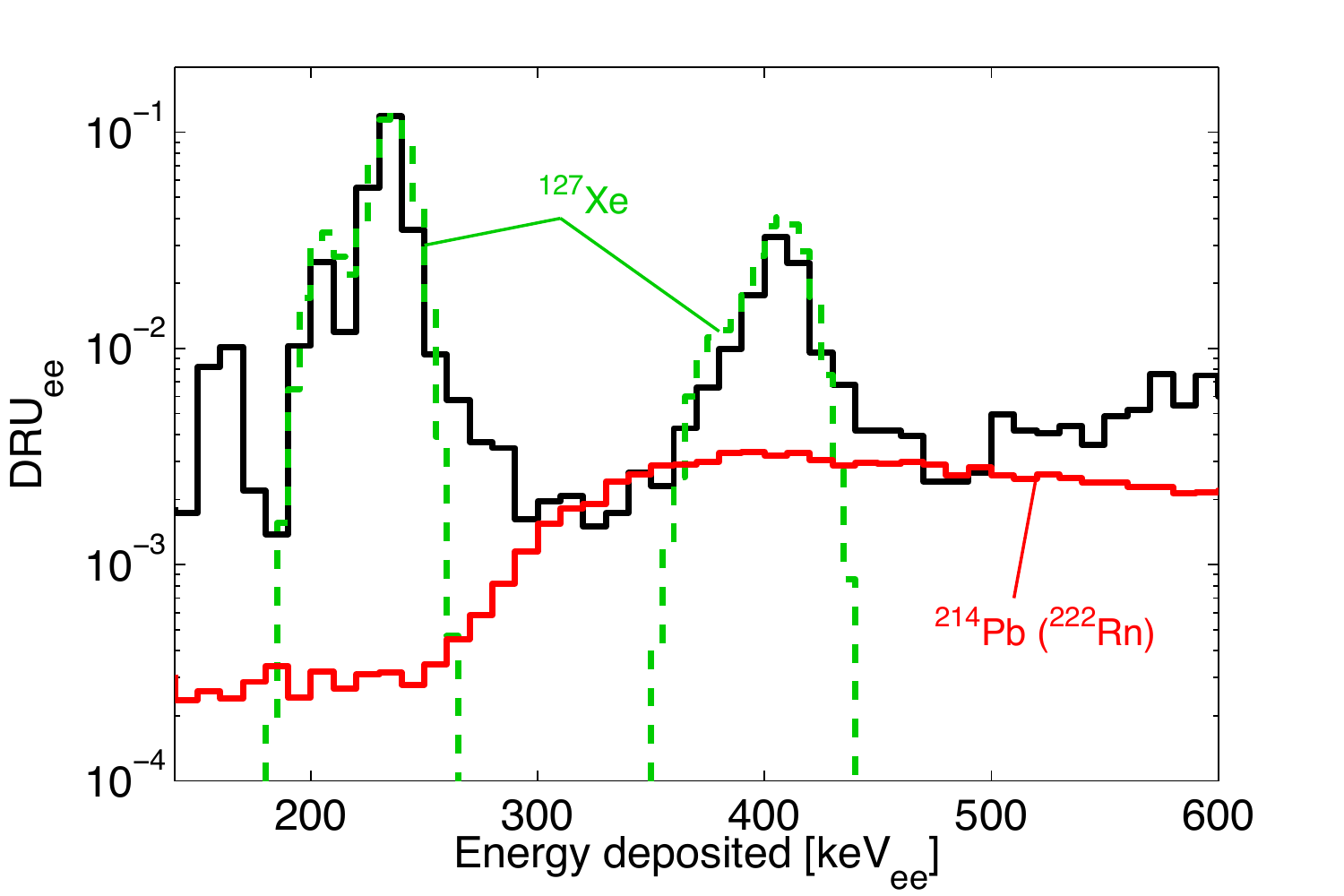}
\par\end{centering}
\caption{\label{fig:Pb214-Constraint}Constraint on $^{214}$Pb in a 30~kg fiducial volume. Spectra are shown for measured data (black), simulation of $^{127}$Xe peaks (gray dashed, green in color), and simulation of $^{214}$Pb (gray solid, red in color). The $^{214}$Pb activity shown is 32~$\mu$Bq~kg$^{-1}$, measured to be the 90\% upper limit from data in the range 300--350~keV$_{ee}$.}
\end{figure}

An additional constraint is placed on the $^{214}$Pb rate from direct measurement of the ER spectrum. The measurement is performed in a low-background 30~kg fiducial volume, where the external $\gamma$~ray depositions are heavily suppressed. The measurement is performed in the range 300--350~keV$_{ee}$, between the peaks from $^{127}$Xe. The measured spectrum, along with models of $^{214}$Pb and $^{127}$Xe spectra, are shown in Fig.~\ref{fig:Pb214-Constraint}. For the purposes of setting a limit on the $^{214}$Pb decay rate, it is conservatively assumed that there is no $\gamma$~ray background in this range and that all activity is due to $^{214}$Pb decay. The $^{214}$Pb decay spectrum is compared with data in this range, using the nearby $^{127}$Xe peaks to calibrate cut efficiencies and estimate energy resolution. The upper limit on $^{214}$Pb activity from this exercise is $<$32~$\mu$Bq~kg$^{-1}$ at 90\% CL, or $<$8~mBq integrated over the entire active region. This exercise provides a much stronger upper limit on the $^{214}$Pb activity than interpolation from $\alpha$~decay rates alone.

The low-energy ER background contribution for $^{214}$Pb is taken from the fraction of its $\beta$~spectrum that falls inside the WIMP search range. This accounts for the ER background generated from naked $\beta$~decay. $^{214}$Pb can also potentially generate a position-dependent background from a semi-naked $\beta$~decay, where the 352~keV $\gamma$~ray escapes from the active region. For the 118~kg fiducial, the naked $\beta$~ray background component comprises $>$95\% of the total background signal. The semi-naked component negligible and is ignored. The total contribution, with a lower bound from the measured $^{214}$Po decay rate and an upper bound from the high-energy ER spectrum, is $0.10-0.22$~mDRU$_{ee}$. Background models assume a concentration of 0.2~mDRU$_{ee}$ $^{214}$Pb, calculated assuming that the $^{222}$Rn chain rates progress as a geometric series, with endpoints constrained by the measured decay rates of the visible $^{222}$Rn daughters.

\subsection{\label{sub:Kr}$^{85}$Kr Removal, Monitoring and Backgrounds}

The research-grade Xe procured for LUX contained an average 130~ppb~g/g $^{\textrm{nat}}$Kr~/~Xe upon acquisition. Natural Kr contains the unstable isotope $^{85}$Kr in estimated concentrations of $2\times10^{-11}$~(g/g) \cite{Kr85Basics}. $^{85}$Kr decays with a half-life of 10.8~years via emission of a $\beta$ with 687~keV endpoint. The $\beta$ emission creates low-energy ER backgrounds at the level of 5~DRU$_{ee}$, at 130~ppb $^{\textrm{nat}}$Kr~/~Xe.

Krypton is not removed by the LUX getter, as the getter removes only non-noble impurities. An independent Kr removal system for LUX Xe was built and operated at Case Western Reserve University. The Kr removal system was established with a goal of reducing $^{85}$Kr background levels to $<$0.2~mDRU$_{ee}$. This level was chosen so that the $^{85}$Kr background would be $\times$0.25 that of the external $\gamma$~ray background projected for a 100~kg fiducial volume. The corresponding $^{\textrm{nat}}$Kr concentration in Xe is $<$5~ppt.

The Kr removal system uses a column of activated charcoal to chromatographically separate Kr from Xe \cite{ABolozKr}. Details of the system will be released in a separate upcoming LUX publication \cite{LUXKrRemoval2014}. From August~2012 to January~2013, a total of 400~kg of Xe was purified with the charcoal system for the first LUX WIMP search run. Krypton was reduced by an average factor $3\times10^4$ for each Xe batch. No other impurities were introduced in significant quantities during the purification process. Loss of Xe during the process was negligible. The final measured Kr concentration in LUX Xe immediately after purification was $4\pm1$~ppt~g/g.

The purified Xe was sampled weekly over the course of the WIMP search run to detect any new Kr signatures. Krypton detection at the ppt level was accomplished using a liquid nitrogen cold trap and mass spectrometry analysis, based on the design presented in \cite{KrDetInXe}. Average Kr levels in LUX Xe during the WIMP search run were measured to be $3.5\pm1.0$~ppt~g/g. The corresponding ER background rate, calculated directly from the $^{85}$Kr beta spectrum, is $0.17\pm0.10$~mDRU$_{ee}$. The error on the background estimate is due to uncertainty in the total Kr concentration in the Xe, as well as uncertainty in the ratio of $^{85}$Kr~/~$^{\textrm{nat}}$Kr, which is taken to be a factor $\times$2.

%%%%%%%%%%%%%%%%%%%%%%%%%%%%%%%%%%%%%%%%%%%%%%%%%%%%%%%%%%%%%%%%%%%%%%%%%%%%%
\subsection{Neutrons from Construction Materials}

\subsubsection{Predicted Yields and LUXSim Modeling}

The dominant source of neutrons in LUX is the $^{238}$U and $^{232}$Th content in the R8778~PMTs \cite{LUXNIM,R11410}. The ($\alpha$,n) neutron spectrum generated from the PMTs was calculated using the Neutron Yield Tool developed by LUX collaborators at the University of South Dakota \cite{NeutronYieldTool,NYTNIM}. The total neutron generation rate is 1.2~n~PMT$^{-1}$~yr$^{-1}$.

An additional contribution to the ($\alpha$,n) neutron background results from $^{210}$Pb that has plated onto detector materials yields. The dominant component of this additional contribution comes from $\alpha$~particle interactions in PTFE, as fluorine generates 1--3~orders of magnitude more neutrons through ($\alpha$,n) reactions than other typical construction materials \cite{ChainDiseqNeutron}. The observed $^{210}$Po 5.3~MeV $\alpha$~decay rate on the PTFE walls of the active region is 14~mHz. The corresponding neutron emission rate, multiplied by $\times$2 to account for PTFE surfaces not visible to the active region, is 8.8~yr$^{-1}$. This emission rate is only 6\% of the total PMT neutron emission rate. Polonium-210 neutrons are not incorporated into the NR background model.

LUXSim was used to simulate the emission of neutrons isotropically from the top and bottom PMT arrays, using the energy spectrum calculated with the Neutron Yield Tool. MS and SS cuts were implemented in post-processing of LUXSim data, mimicking the cuts used in WIMP search data. Neutron events were selected in the range 3.4--25~keV$_{nr}$. The total number of SS neutron events predicted in the WIMP search energy range in 85.3~days $\times$ 118~kg fiducial volume is 0.06.

\subsubsection{\label{sub:Neutron-MS}Multiple Scatter Identification in LUX Data}

Radiogenic neutron SS events mimic a WIMP signature in both NR energy spectrum and ER/NR discrimination, as described in Sec.~\ref{sec:Introduction}. For this reason, and because the neutron SS rate is expected to be very low in LUX, a direct search for neutron SS events cannot be used to place meaningful constraints on the neutron background. The 47~cm LUX diameter creates an environment in which the number of neutron MS events are much greater than the number of neutron SS events, due to the $\lesssim$10~cm mean free path of typical radiogenic neutrons. Constraints on neutron MS event rates were used to place an upper limit on the neutron SS event rates.

Simulated neutron background studies were used to calculate the ratio of the number of NR MS events within an enlarged 180~kg fiducial volume to the number of SS events within a smaller WIMP search fiducial volume. The 180~kg fiducial is chosen as the maximum volume where low-energy MS events are reliably reconstructed. WIMP search fiducial volumes with masses 118~kg and 100~kg are explored, where the 118~kg volume corresponds to the fiducial volume used for the 85.3~day WIMP search run, and the 100~kg volume is a nominal LUX one-year WIMP search run fiducial chosen for consistency with previous background studies. LUXSim simulations estimate a ratio of 13 MS events in the 180~kg fiducial per SS events in the 118~kg fiducial volume.

The MS NR and ER S2/S1 discrimination bands were constructed using NEST and measured LUX efficiencies for S1 and S2 detection \cite{LUXPRL}. The bands were used to characterize the observed MS events as ER-like or NR-like. The S1 and S2 signals for MS events were defined as the summed, position-corrected S1 and S2 signals from all scattering vertices. Construction of the ER MS band used the assumption of a flat-energy, two-scatter population, corresponding to the expected MS background from $\gamma$~ray sources. Construction of the NR MS band used the MS energy depositions from LUXSim neutron background studies.

\begin{figure}[htp]
\begin{centering}
\includegraphics[width=1\columnwidth]{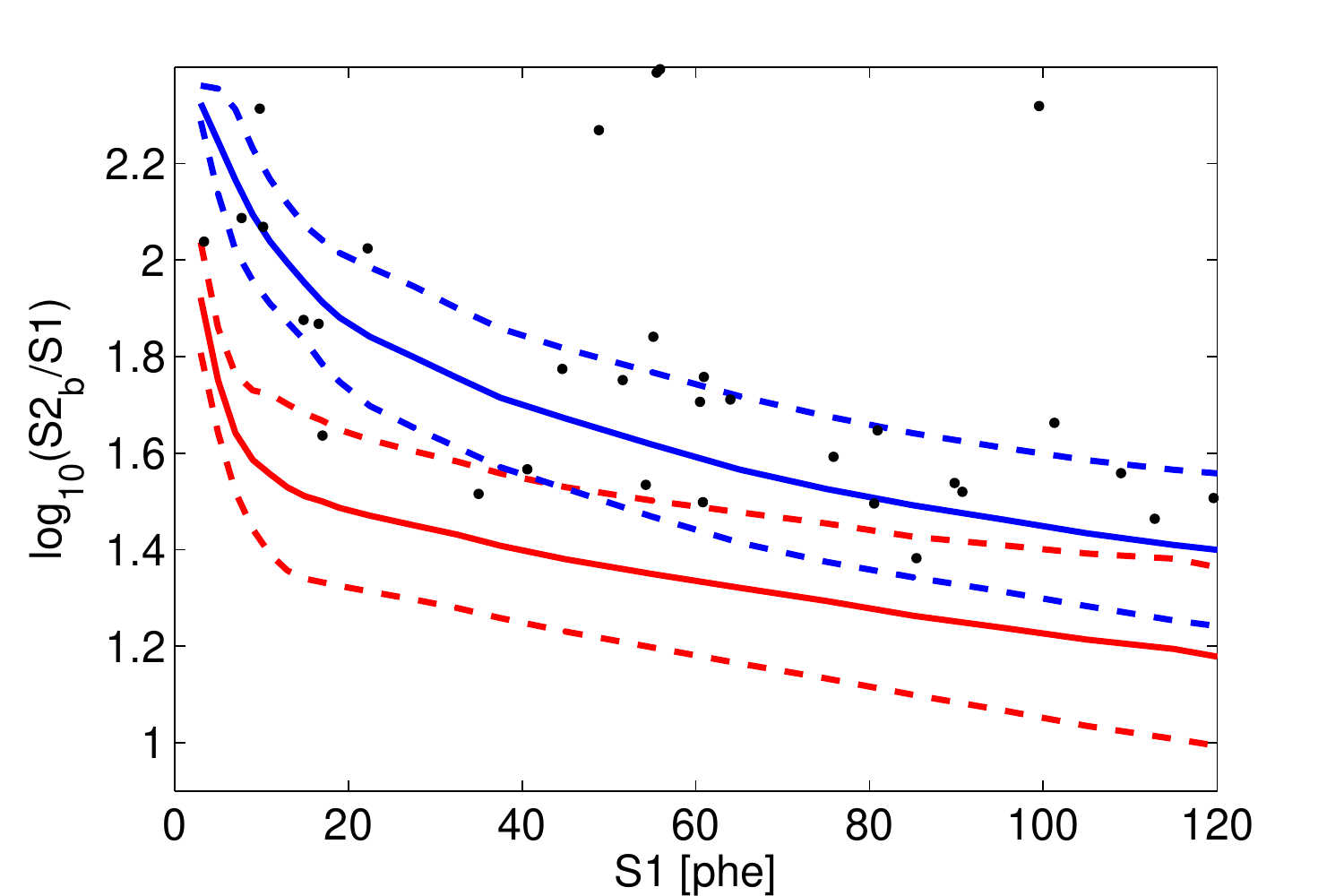}
\par\end{centering}
\caption{\label{fig:Neutron-MS-S2S1-S1}Low-energy MS events in LUX during the 85.3~day WIMP search run, within a 180~kg fiducial volume. Events are plotted in ER/NR discrimination space.  The subscript "b" in S2$_b$ denotes S2 signals from the bottom PMT array only, chosen to avoid irregularities in S2 signals due to deactivated PMTs in the top array. The MS events are used to place an upper limit on the NR SS rate for the 85.3~day and one-year WIMP search runs. Overlaid are the projected bands for ER (black, blue in color) and NR (gray, red in color) MS events, shown with centroids (solid) and $\pm1.28\sigma$ bounds (dashed). Events well above the ER band are found to be inconsistent with MS events in the search volume.}
\end{figure}

MS selection cuts mirror the cuts used to select single-scatter NR events in WIMP search data, but requiring events with $>$1~S2 pulse, each with $>$200~phe, and within a S1 range of 0--120~phe. The cuts are estimated to have a 95\% efficiency in catching all NR MS from the predicted neutron energy spectrum. Candidate MS events are shown along with the $\pm$1.28$\sigma$ ER and NR bands in Fig.~\ref{fig:Neutron-MS-S2S1-S1}. The NR MS search region was defined below the NR band centroid, giving an overall search efficiency of 48\%. No NR-like MS events were observed in the 180~kg search region in 85.3~days, corresponding to an upper-limit of 2.3 MS events at 90\% CL. A population of events well above both the ER and NR bands are found, which are comprised of events in the gas Xe region, and other events with waveform topologies generally inconsistent with MS events in the 180~kg search region.

Given the ratio of 13~MS to SS events, and the search efficiency of 48\%, an upper-limit of 2.3 NR MS events sets a NR SS event upper limit for the run of $<$0.37~events in the 118~kg fiducial volume. The expectation based on Monte Carlo results is 0.06~events. For a one-year run using a 100~kg fiducial, the upper limit is $<$0.72~events, with an expectation of 0.1.

%%%%%%%%%%%%%%%%%%%%%%%%%%%%%%%%%%%%%%%%%%%%%%%%%%%%%%%%%%%%%%%%%%%%%%%%%%%%%
\subsection{External Backgrounds}

\subsubsection{Water Shield Design and Radiogenic Cavern Backgrounds}

The LUX detector is placed inside a 300~tonne water shield designed to render all external backgrounds subdominant to internal backgrounds. The design of the water shield was assisted by Monte Carlo simulations of both $\gamma$~ray and neutron external backgrounds. The primary factor driving the size and configuration of the water shield was the reduction of the high-energy muon-induced neutron background to the level of 0.1~WIMP-like events in 100~kg $\times$ one-year \cite{LUXNIM}. The cylindrical water shield is 7.6~m in diameter and 6.1~m in height, providing a minimum water thickness of 2.75~m at the top, 3.5~m on the sides, and 1.2~m on the bottom. It is built on top of 20~tonnes of low-radioactivity steel plates arranged in an inverse pyramid configuration, with a maximum thickness of 31~cm. The shape of the inverse steel pyramid is optimized to reduce the $\gamma$~ray flux originating from the rock below, multiplying the total external $\gamma$~ray rate in the detector by $\times$1/40 \cite{deViveiros2009thesis}.

The external $\gamma$~ray background is dominated by decays of $^{40}$K and the $^{238}$U and $^{232}$Th chains in the cavern rock. Radiometric surveys of the Homestake mine indicate that most rock in the 4850~ft level are of the type labeled HST-06, consisting of 0.160 ppm $^{238}$U, 0.200 ppm $^{232}$Th and 1540 ppm $^{40}$K \cite{HomestakeRock:2007}. Geological surveys also show rhyolite intrusions in the rock with much higher radioactivities, with average contamination levels of 8.6 ppm $^{238}$U, 10.8 ppm $^{232}$Th, and 29000 ppm $^{40}$K. The percentage of rhyolite intrusions on the cavern surface is unknown, and in order to set a conservative estimate of the $\gamma$~ray event rate in the detector, background estimates assume a cavern completely composed of rhyolite, resulting in a flux of 9~$\gamma$~cm$^{-2}$~s$^{-1}$ at the water shield outer surface. Radioactive screening of typical concrete mixes indicate that the radioactive contamination levels in concrete are in the range of 1--2~ppm and are well below the conservative assumption for the surrounding rock \cite{Wulandari2004,NCRPreport94}.

Monte Carlo simulations of the water shield use a ``standard rock'' $\gamma$~ray energy spectrum obtained from measurements at the Boulby Mine \cite{Carson2005418} and scaled to match the radioactivity levels assumed for the Davis cavern. The simulations show that $\gamma$~ray flux is reduced by a factor of $2\times10^{-10}$ by the water shield and steel pyramid. The resulting $\gamma$~ray flux incident on the LUX detector generates a low-energy ER event rate of 27~nDRU$_{ee}$ in the active region.

The external neutron flux is dominated by the environmental fast neutron background (E~>~1~MeV) due to radioactive processes in the surrounding rock and concrete. The neutrons are mainly produced by $^{238}$U spontaneous fission and ($\alpha$,n) neutrons generated in the rock and concrete from the $^{238}$U and $^{232}$Th chain decays. Although there are no published measurements of the neutron flux in the Davis laboratory, the environmental neutron background can be estimated by scaling the flux measured in the Gran Sasso laboratory \cite{Wulandari2004} to match the contamination levels found in the Homestake rock. Using the conservative limit that the Davis cavern is completely lined by rhyolite rock, the estimated incident neutron flux is $16.2\times10^{-9}$~n~cm$^{-2}$~s$^{-1}$. The fast neutron flux is efficiently moderated by the water shield, and the total integrated flux reduction due to water plus geometry is $6\times10^{-22}$ for E~>~1~keV. This corresponds to $10^{-16}$~n~yr$^{-1}$ incident on the detector.

\subsubsection{Muon-Induced Neutron Backgrounds}

Cosmic ray muons contribute to the NR background through the generation of neutrons both in the shield and surrounding rock. The resulting neutron flux has a significant high-energy component (E~>~10~MeV), reduced by only $\sim$3 orders of magnitude in the water shield \cite{LUXNIM}. The choice of a deep site is essential in controlling this neutron background.

The Davis laboratory is located in the 4850~ft level of the Homestake mine, with an effective depth of $4.3\pm0.2$~km.w.e. This depth corresponds to a muon flux of $\left(4.4\pm0.1\right)\times10^{-9}$~cm$^{-2}$~s$^{-1}$ \cite{Mei:2005gm} and an average muon energy of 321~GeV \cite{2001ADNDT..78..183G}. The muon-induced high-energy neutron flux from the rock has been measured at several underground sites, and both the total flux and energy distribution can be fitted with empirical depth-dependent functions \cite{Mei:2005gm}. The neutron flux from the rock at Homestake is calculated to be $\left(0.54\pm0.01\right)\times10^{-9}$~n~cm$^{-2}$~s$^{-1}$, where the quoted error reflects simulation statistics only and does not reflect the full uncertainty due to local variations in rock content. The energy distribution is estimated by using the function fit parameters for the Gran Sasso laboratory, which is the closest possible site in depth to the Homestake 4850~ft level for which neutron flux from rock has been measured. The water shield reduces the integrated flux of high-energy rock neutrons at the cryostat to $1\times10^{-7}$~n~s$^{-1}$, resulting in a SS NR rate of 60~nDRU$_{nr}$ in the $3.4-25$~keV$_{nr}$ range, for a 100~kg fiducial volume.

The muon-induced neutron production in the water shield can be estimated from the neutron yield in different materials, calculated through Monte Carlo simulations using MACRO and Geant4 \cite{Mei:2005gm}. For a water tank of 7.6~m diameter and 6.1~m height, the total neutron flux at the LUX outer cryostat is $6.3\times10^{-7}$~n~s$^{-1}$. The resulting NR event rate in the 100~kg fiducial region is 120~nDRU$_{nr}$ in the energy range of 3.4--25~keV$_{nr}$.

The total nuclear recoil background due to muon-induced high-energy neutrons, including the components from the cavern rock and generated in the water itself, is 0.1~WIMP-like NR events in the energy range of 3.4--25~keV$_{nr}$, in 100~kg $\times$ one-year. The projected rate meets the external neutron background goal. External backgrounds are not considered for the 85.3~day WIMP search run, and are listed in the summary for the one-year run in Table~\ref{tab:BG-1yr}.

%%%%%%%%%%%%%%%%%%%%%%%%%%%%%%%%%%%%%%%%%%%%%%%%%%%%%%%%%%%%%%%%%%%%%%%%%%%%%
\section{\label{sec:ER-NR-Disc}ER/NR Discrimination}

LUX relies on differences in the ionization~to~scintillation ratio between ER and NR events to provide rejection of low-energy ER events falling inside the WIMP search fiducial volume. The discrimination power is mapped through a combination of LUXSim/NEST Monte Carlo studies and direct calibration measurements.

\begin{figure}
\begin{centering}
\subfloat[]{\includegraphics[width=1\columnwidth]{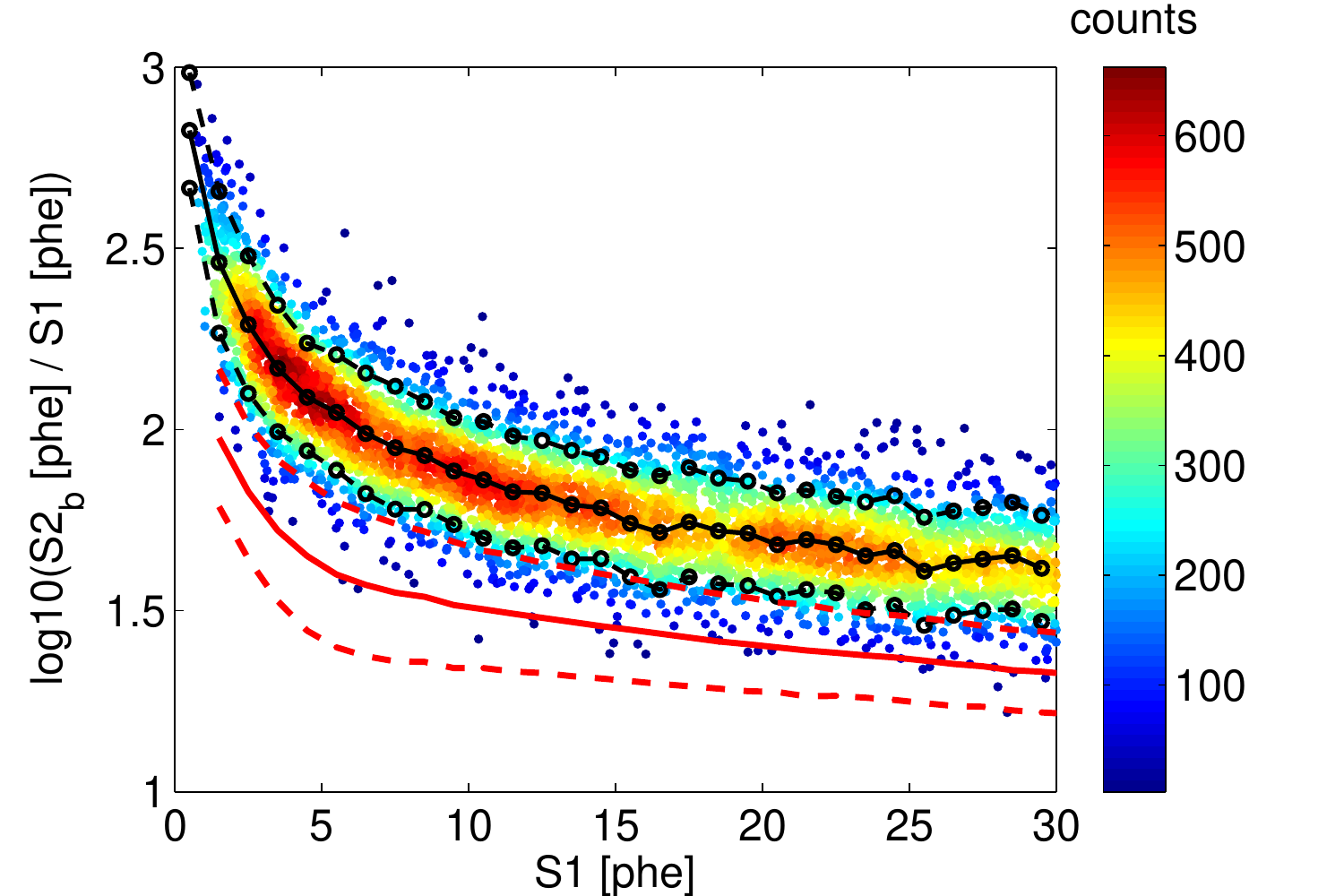}
\label{fig:ER-Band}}

\subfloat[]{\includegraphics[width=1\columnwidth]{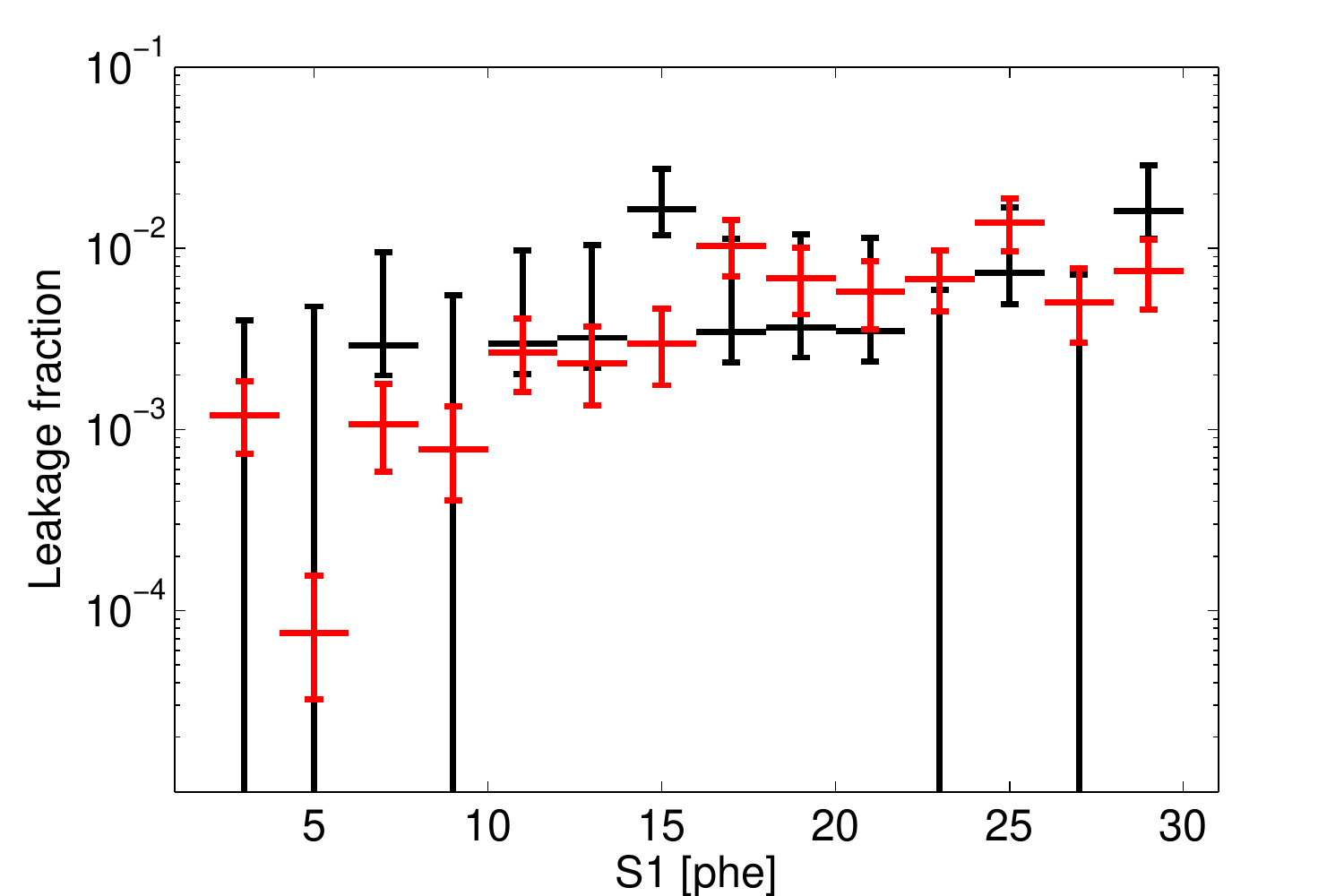}
\label{fig:Tritium-ER-Leakage}}
\par\end{centering}
\caption{\label{fig:ER-Band-and-Leakage}(a) ER S2/S1 band in LUX, as measured by $^3$H calibration. The $^3$H run yielded 4400~decays in the 118~kg fiducial volume, and are used to map the band mean and width. The measured band centroid is overlaid (solid), with $\pm$1.28$\sigma$ contours (dashed). The NR band, calculated using NEST and verified by neutron calibration, is shown in red. (b) Measurement of ER band leakage below the NR band centroid for the LUX 85.3~day WIMP search run. Points are shown corresponding to the measured leakage fraction in each S1 bin (black), and the projected leakage fraction based on a Gaussian fit to the ER band in each bin (gray, red in color). Data is volume-averaged over the entire 118~kg fiducial volume. Measured points are taken from $^3$H calibration. Errors shown are $\pm$34.1\%.}
\end{figure}

The position and width of the ER S2/S1 band was mapped in the WIMP search range using $10^4$~$^3$H decays \cite{LUXPRL}. The $^3$H was delivered into the active region by injection of tritiated methane, using a system developed by LUX collaborators at the University of Maryland \cite{Tritium}. The system will be described in detail in a future LUX publication \cite{LUXTritium2014}. Tritium offers a unique ability to map detector ER response in the WIMP search range, decaying by a naked $\beta$ with endpoint 18.6~keV. 45\% of $^3$H decays fall within the 0.9--5.3~keV$_{ee}$ LUX WIMP search energy range. The measured ER band $\pm$1.28$\sigma$ as a function of S1 photoelectrons is shown in Fig.~\ref{fig:ER-Band}.

The NR band (also shown on the same figure) shape was generated using LUXSim NR events, convolved with both measured LUX S1 and S2 detection efficiencies and NEST photon and electron distributions. Direct NR calibration through AmBe and $^{252}$Cf sources was used to verify the simulated NR band mean and width. These sources did not provide sufficient statistics for detailed NR band mapping, due to the large detector volume and the low fraction of SS NR events achievable with external calibration sources. The NR calibration program will be described in detail in a future LUX publication \cite{LUXNR2014}.

The measured ER leakage fraction is shown in Fig.~\ref{fig:Tritium-ER-Leakage}. The ER leakage fraction is defined as the fraction of ER events falling below the NR band centroid in a given S1 bin. Note that this ER discrimination definition corresponds to a "cut-and-count" WIMP search analysis, analogous to that performed for previous Xe dark matter experiments \cite{XENON10,ZEPLINIII}. LUX dark matter search results from the 85.3~day run used a profile likelihood ratio (PLR) analysis that extended the dark matter event search above the NR band centroid \cite{LUXPRL}. The ER leakage fraction is used as a conservative expectation of the ER background rejection power for the detector.

The leakage fraction is calculated both from direct measurement in $^3$H calibration, and from a Gaussian fit to the ER band. The average leakage fraction in the WIMP search energy range 2--30~phe, for both direct measurement and Gaussian fits, is 0.04. This corresponds to 99.6\% ER discrimination.

%%%%%%%%%%%%%%%%%%%%%%%%%%%%%%%%%%%%%%%%%%%%%%%%%%%%%%%%%%%%%%%%%%%%%%%%%%%%%
\section{\label{sec:Comparison-with-lowE-data}Comparison with Measured Low-Energy Data}

The predicted WIMP search ER background from the sources listed in Sec.~\ref{sec:Background-Modeling} is compared with measured LUX background data from the 85.3~day WIMP search run. The WIMP search fiducial volume is used, defined as a cylinder with a radius of 18~cm and a height of 40~cm. The fiducial volume is centered radially in the detector, with 7~cm Xe (1.4~cm in the drift region) below and 7~cm (6.5~cm in the drift region) above. Backgrounds are evaluated over the WIMP search energy range, 2--30~phe (0.9--5.3~keV$_{ee}$), in order to encompass the $^{127}$Xe spectral shape. All other background spectra are flat in energy, and do not change the differential background measurement.

\begin{figure*}
\begin{centering}
\subfloat[]{\includegraphics[width=0.5\textwidth]{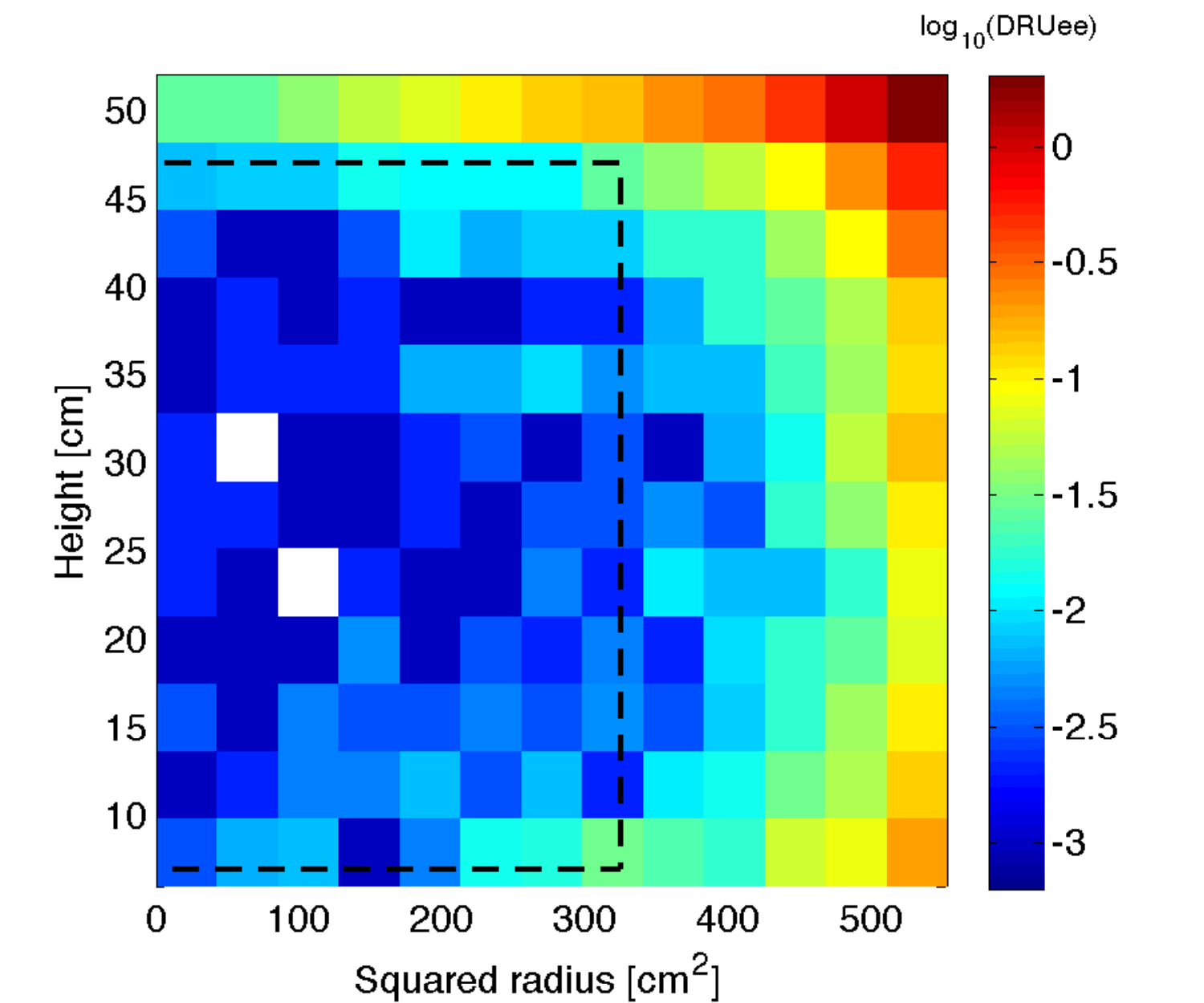}
}
\subfloat[]{\includegraphics[width=0.5\textwidth]{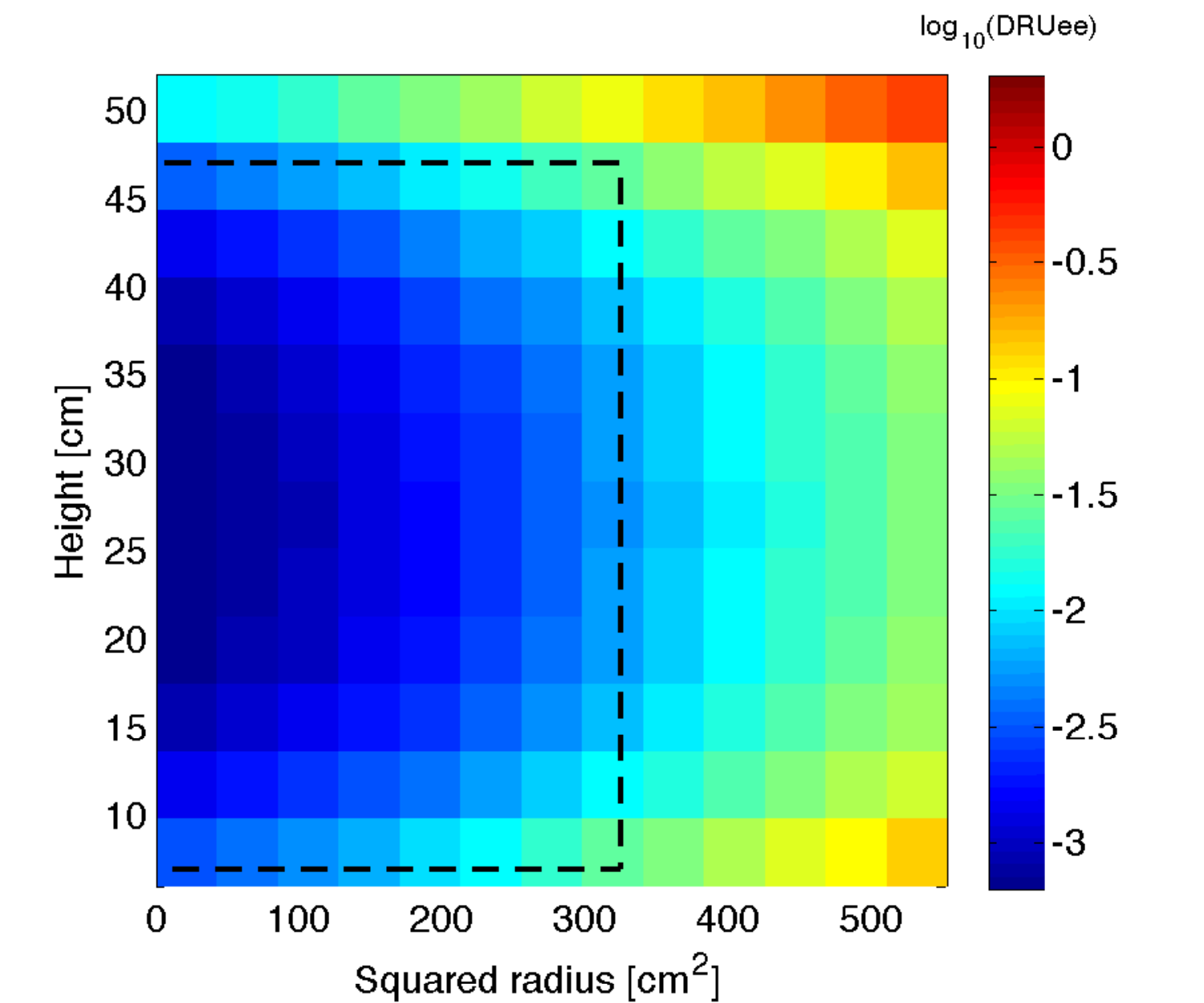}
}
\par\end{centering}
\caption{\label{fig:BG-RZ-dist}Low-energy background distributions in squared radius and height, from (a) measured data and (b) model predictions. Rates are taken in the range 0.9--5.3~keV$_{ee}$ (2--30~S1~phe). Rates are shown in units of $\log_{10}\left(\textrm{DRU}_{ee}\right)$. The 118~kg fiducial volume used in the 85.3~day WIMP search run is shown in dashed black. The model includes low-energy background contributions from $\gamma$~ray, $^{127}$Xe, $^{214}$Pb, and $^{85}$Kr sources. Measured rates at large radii include a significant contribution from low-energy $^{210}$Pb decays at the detector walls. These decays are not included in the background model.}
\end{figure*}

\begin{figure}
\begin{centering}
\subfloat[]{\includegraphics[width=0.95\columnwidth]{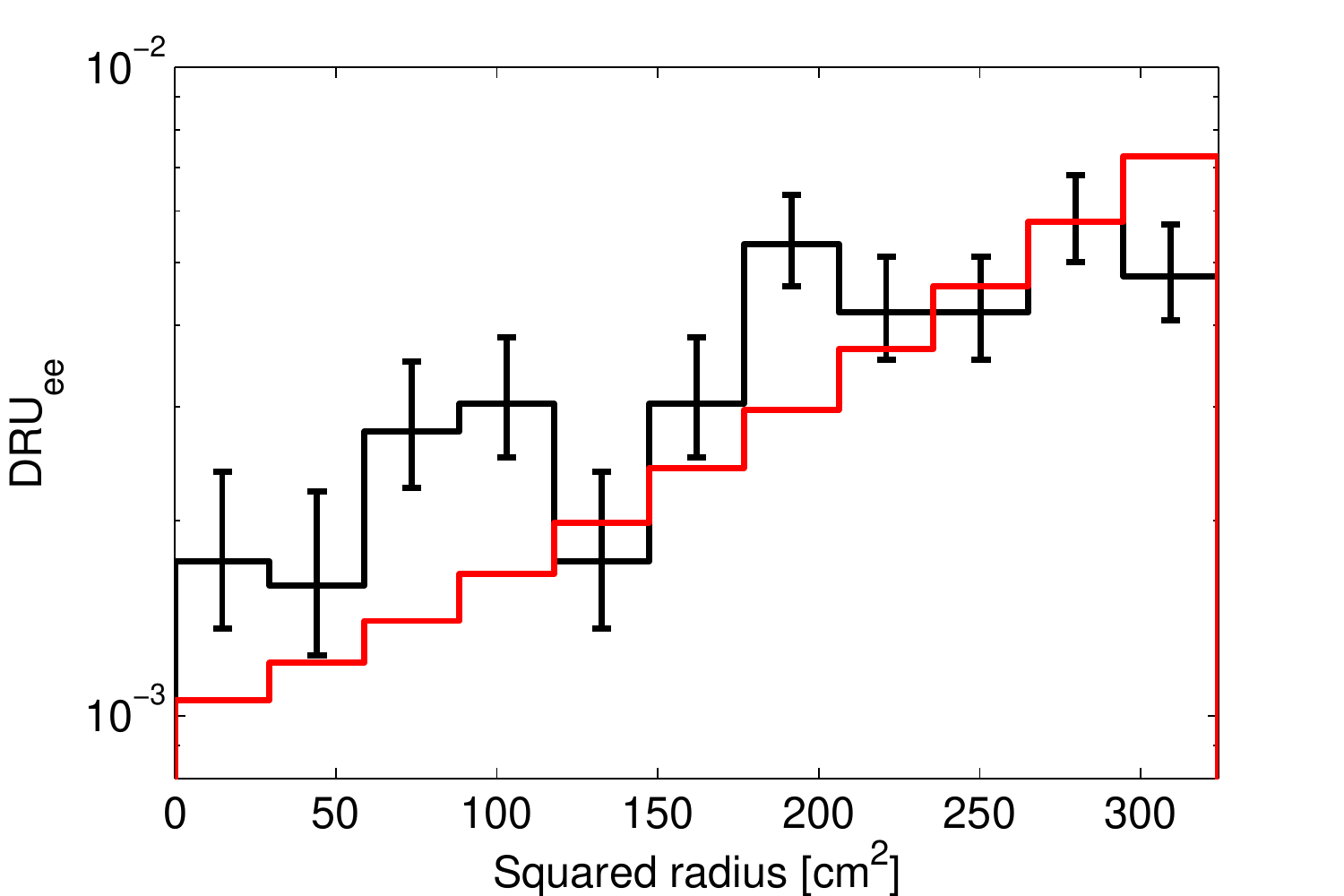}
}

\subfloat[]{\includegraphics[width=0.95\columnwidth]{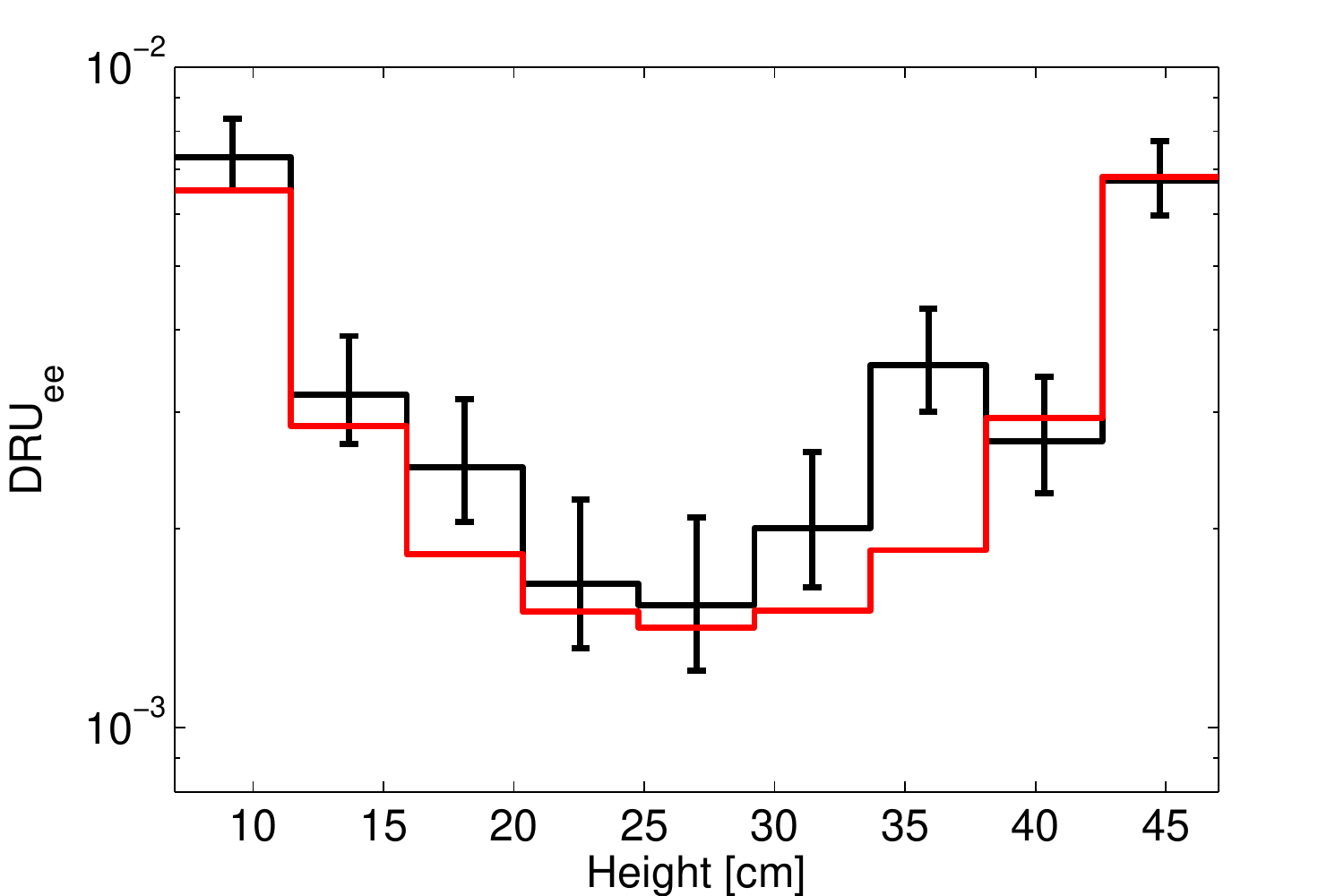}
}

\subfloat[]{\includegraphics[width=0.95\columnwidth]{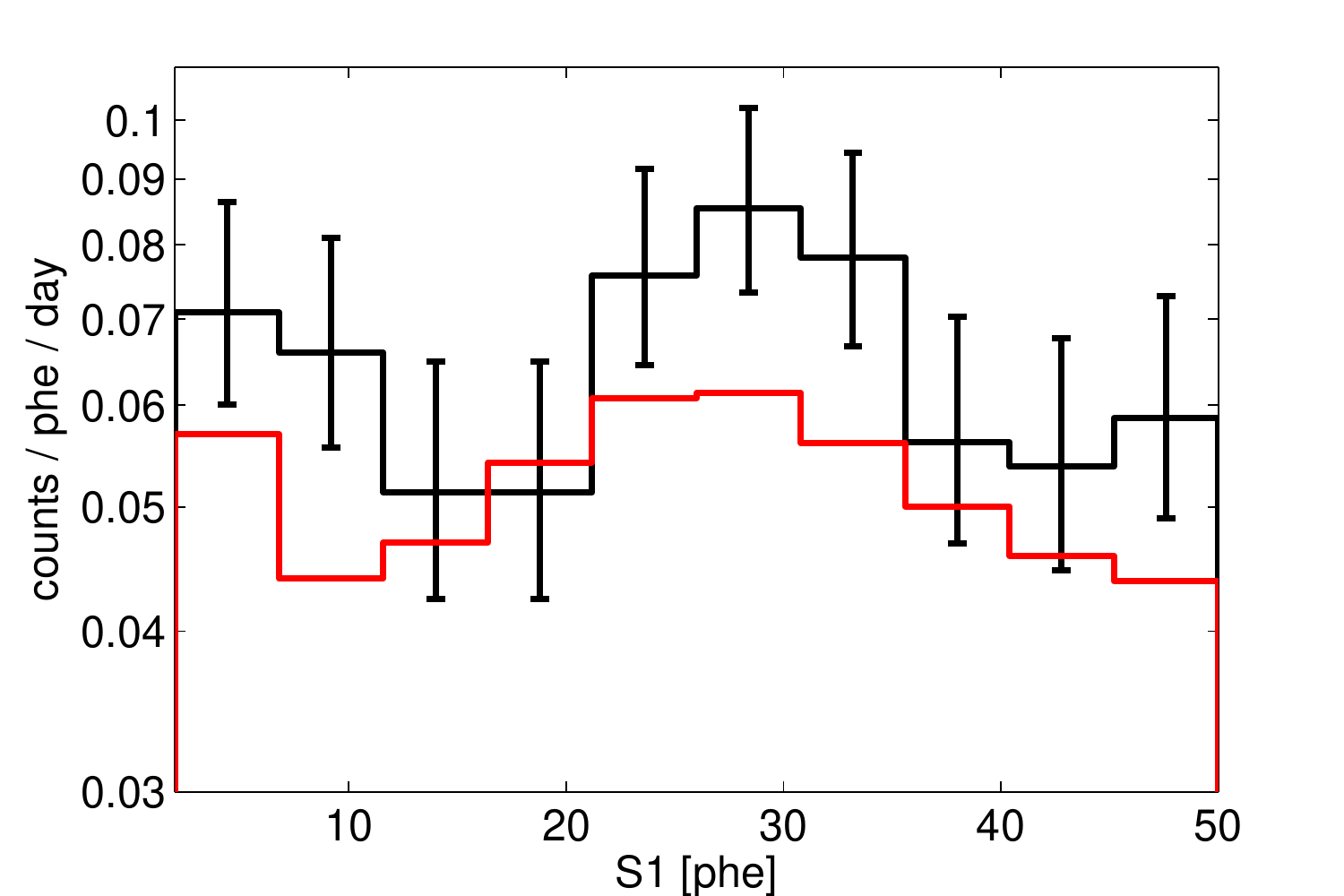}
}
\par\end{centering}
\caption{\label{fig:BG-dist-meas}Low-energy measured distributions in (a) squared radius, (b) height, and (c) S1, within the LUX 118~kg fiducial volume, measured over the full 85.3~day WIMP search run. Measured data are indicated by the black histogram with error bars. Simulation data are shown as the gray histogram (red, in color). Simulated radial and height distributions are reconstructed from high-energy background studies, and are not a fit to low energy distributions. The simulated S1 distribution folds in NEST estimates of total photon yields and measured LUX light collection efficiency.}
\end{figure}

The measured and simulated background rates as a function of position in the detector are shown in Fig.~\ref{fig:BG-RZ-dist}. The flattened radial, height and S1 distributions in the WIMP search fiducial volume for simulation and measured data are shown in Fig.~\ref{fig:BG-dist-meas}. The S1 spectrum is constructed based on LUXSim background studies, NEST photon and electron yields, and measured LUX S1 and S2 detection efficiencies, using the same technique performed for construction of the NR band in Sec.~\ref{sec:ER-NR-Disc}.

Use of Kolmogorov-Smirnov (K-S) tests for the height and S1 distribution shapes yields p~values of 26\% and 94\% respectively when testing the measured data against the simulated distributions. The radial distribution is measured to be systematically flatter than simulation predictions, with a K-S test p~value of 0.004\%. The background expectation averaged over the entire fiducial volume and WIMP search run is given in Table~\ref{tab:Predicted-measured-backgrounds}.

\begin{figure}
\begin{centering}
\includegraphics[width=1\columnwidth]{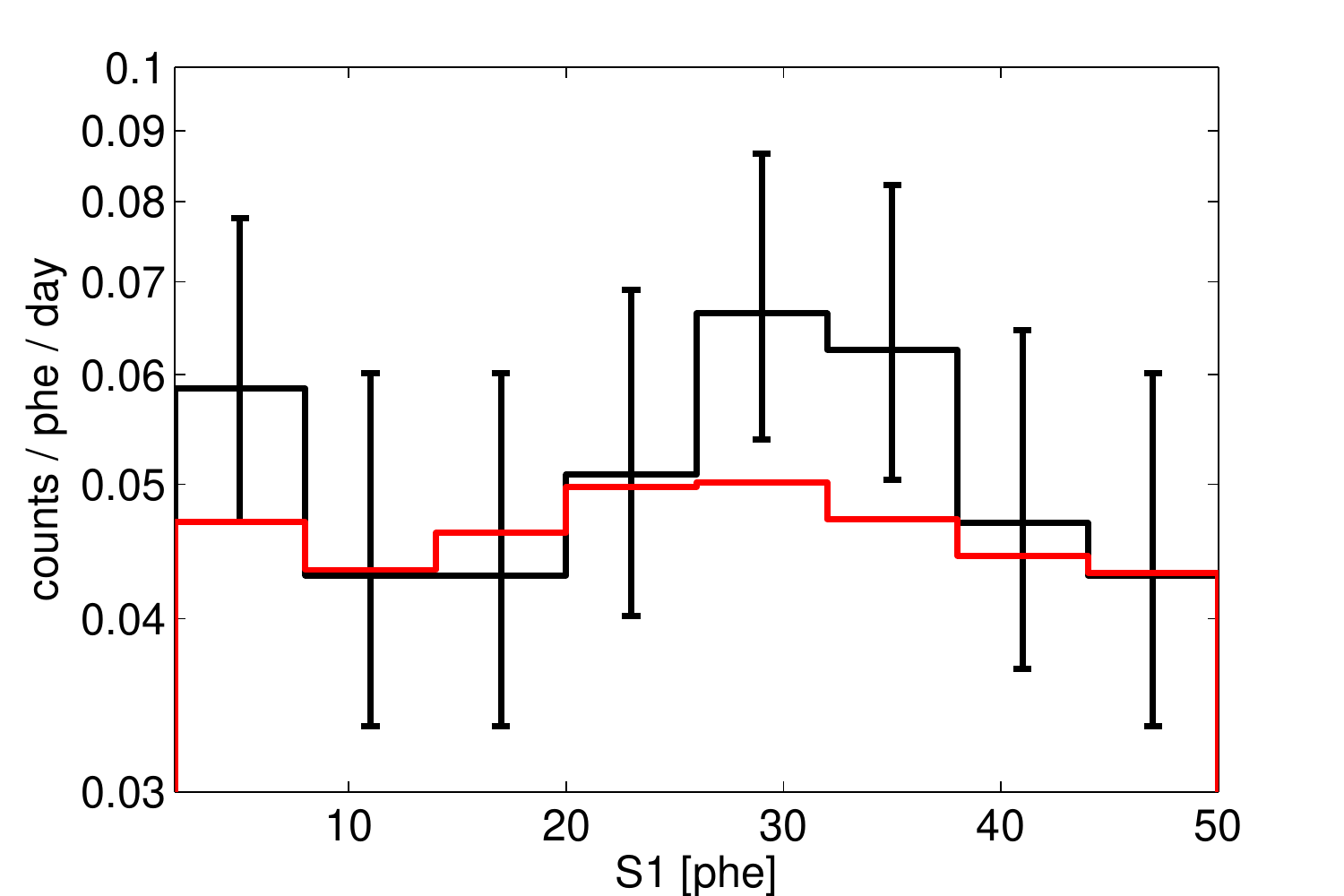}
\par\end{centering}
\caption{\label{fig:S1-BG-Dist-Late}Low-energy ER backgrounds in the 118~kg fiducial volume during the second half of the 85.3~day WIMP search run (after June~14). Measured data are shown in black. The model predictions, including $\gamma$~ray, $^{127}$Xe, $^{214}$Pb, and $^{85}$Kr, are shown in gray (red, in color). Model S1 predictions are based on measured LUX efficiency factors and NEST photon and electron distributions. The measured data totals $2.8\pm0.4_{\textrm{stat}}$~mDRU$_{ee}$.}
\end{figure}

WIMP search ER backgrounds in the range 0.9--5.3~keV$_{ee}$ in the fiducial volume are shown after the June~14 midpoint of the WIMP search run in Fig.~\ref{fig:S1-BG-Dist-Late}. The fiducial ER background rate in the first half of the run is $4.4\pm0.4_{\textrm{stat}}$~mDRU$_{ee}$, while the background rate in the second half is $2.8\pm0.4_{\textrm{stat}}$~mDRU$_{ee}$. The rate drop is a factor $\times$2.7 higher than that predicted due to the decay of cosmogenic $^{127}$Xe alone. However, the background rate in the second half of the run is consistent with the predicted $2.2\pm0.3$~mDRU$_{ee}$ due to time-independent sources ($^{214}$Pb, $^{85}$Kr, and $\gamma$~rays from construction materials), with an additional $0.28\pm0.06$~mDRU$_{ee}$ from $^{127}$Xe. A K-S test yields a p~value of 87\% from comparison of the measured and simulated population distribution shapes.

The observed event distribution in S2/S1 is consistent with an ER population, with total rate matching predictions based on modeling work \cite{LUXPRL}. From the measured 99.6\% discrimination factor, an average 0.64~events are expected to fall below the NR band centroid. One~event is observed at the NR centroid, at 3~phe~S1. The PLR WIMP search analysis finds agreement with the background-only hypothesis with a p~value of 35\%.

\begin{table}
\begin{centering}
\begin{tabular}{ccccc}
\hline 
\textbf{Source}		& \textbf{Background Rate [mDRU$_{ee}$]} \tabularnewline
\hline
\hline
$\gamma$~rays	& $1.8\pm0.2_{\textrm{stat}}\pm0.3_{\textrm{sys}}$ \tabularnewline
$^{127}$Xe		& $0.5\pm0.02_{\textrm{stat}}\pm0.1_{\textrm{sys}}$ \tabularnewline
$^{214}$Pb		& $0.11-0.22$ (0.20 expected) \tabularnewline
$^{85}$Kr			& $0.17\pm0.10_{\textrm{sys}}$ \tabularnewline
\hline
\hline
Total predicted		& $2.6\pm0.2_{\textrm{stat}}\pm0.4_{\textrm{sys}}$ \tabularnewline
Total observed		& $3.6\pm0.3_{\textrm{stat}}$ \tabularnewline
\hline
\end{tabular}
\par\end{centering}
\caption{\label{tab:Predicted-measured-backgrounds}Predicted and measured low-energy background rates in the LUX 118~kg WIMP search fiducial during the 85.3~day run. Rates are averaged over the energy range 0.9--5.3~keV$_{ee}$.}
\end{table}

%%%%%%%%%%%%%%%%%%%%%%%%%%%%%%%%%%%%%%%%%%%%%%%%%%%%%%%%%%%%%%%%%%%%%%%%%%%%%
\section{Background Projections for the One~Year Run}

The background studies from the 85.3~day WIMP search can be used to project the expected backgrounds for the 2014 one-year LUX WIMP search run. At the beginning of the one-year run, the $^{127}$Xe background will have decayed below significance. The one-year run is also expected to use a more conservative 100~kg fiducial volume, further reducing position-dependent $\gamma$~ray backgrounds.

The predicted background sources within the 100~kg fiducial for the one-year run are listed in Table~\ref{tab:BG-1yr}. A total of $1.4\pm0.2$~mDRU$_{ee}$ is expected from all ER sources, assuming no change in $^{214}$Pb or $^{85}$Kr rates from those observed in the 85.3~day run. The predicted total is in agreement with observations of data in a 100~kg fiducial during the second half of the 85.3~day run. The observed event rate is $1.7\pm0.3$~mDRU$_{ee}$. The observed rate includes $0.15\pm0.04$~mDRU$_{ee}$ of residual $^{127}$Xe, which will not be present during the one-year run. The neutron differential rate from both internal and external sources is 350~nDRU$_{nr}$.

Integrating over the 0.9--5.3~keV$_{ee}$ window for the ER sources and the equivalent 22~keV$_{nr}$ window for NR sources, and using the observed $1.7\pm0.3$~mDRU$_{ee}$ and subtracting $0.15\pm0.04$~mDRU$_{ee}$ $^{127}$Xe, the total expected number of background events is 250 (ER) + 0.28 (NR). After 99.6\% ER discrimination, and assuming a 50\% NR acceptance, the number of WIMP-like background events is $1.1\pm0.2$. The background rate is potentially further reduced by optimizing the shape of the fiducial volume to follow the background contours in the active region. The optimal shape will be determined by observed background rates before the start of the one-year run.

\begin{table}
\begin{centering}
\begin{tabular}{ccccc}
\hline 
\textbf{Source} & \textbf{Background Rate} \tabularnewline
\hline
\hline
$\gamma$~rays	& $\left(1.0\pm0.1_{\textrm{stat}}\pm0.1_{\textrm{sys}}\right)$~mDRU$_{ee}$ \tabularnewline
$^{214}$Pb		& 0.2~mDRU$_{ee}$ \tabularnewline
$^{85}$Kr			& $\left(0.17\pm0.10_{\textrm{sys}}\right)$~mDRU$_{ee}$ \tabularnewline
Int. neutrons	& 170~nDRU$_{nr}$ \tabularnewline
Ext. neutrons	& 180~nDRU$_{nr}$ \tabularnewline
\hline
\hline
Total predicted		& $1.4\pm0.2$~mDRU$_{ee}$ + 350~nDRU$_{nr}$ \tabularnewline
\hline
Total observed		& $1.7\pm0.3$~mDRU$_{ee}$ ($0.14\pm0.03$ $^{127}$Xe) \tabularnewline
\hline
\end{tabular}
\par\end{centering}
\caption{\label{tab:BG-1yr}Predicted and measured low-energy background rates in a 100~kg WIMP search fiducial expected to be used for the one-year run. ER rates are averaged over the energy range 0.9--5.3~keV$_{ee}$. NR rates are averaged over the energy range 3.4--25~keV$_{nr}$. Measured rates are taken from the second half of the 85.3~day WIMP search run. The $^{127}$Xe contribution to the observed background rate is given in brackets. This component will not be present during the one-year run.}
\end{table}

%%%%%%%%%%%%%%%%%%%%%%%%%%%%%%%%%%%%%%%%%%%%%%%%%%%%%%%%%%%%%%%%%%%%%%%%%%%%%
\section{Conclusions}

ER and NR low-energy backgrounds in the LUX experiment have been modeled in detail. Modeling work is based on Monte Carlo projections constrained by $\gamma$~ray assay of construction materials, as well as {\it in-situ} measurements of $\gamma$~rays and intrinsic radioisotope decay rates performed outside of the WIMP search fiducial volume and energy range. Low-energy background predictions are not directly fit from Monte Carlo but rather extrapolated from high-energy measurements. The use of independent measurements to set the model parameters and the resulting good agreement between low-energy projections and observed data gives high confidence that the low-energy backgrounds in LUX are well understood.

The primary backgrounds in the LUX detector arise from low-energy depositions from $\gamma$~ray scatters in the fiducial region. The $\gamma$-rays are generated from radioisotope decays in detector construction materials. The R8778~PMTs are the largest source of $\gamma$~ray backgrounds, with additional contributions from insulation materials. Cosmogenic production of $^{60}$Co in Cu contributes a $\gamma$~ray rate $\times$3 higher than expected based on initial screening results.

Measurements of $\alpha$~particle energy depositions in the detector provide a model for radon daughter decays in the fiducial volume. Alpha decay rates, combined with high-energy spectrum measurements, provide a constraint on $^{214}$Pb rates within a factor of $\times$2. $^{85}$Kr backgrounds are calculated from direct measurements of $^{\textrm{nat}}$Kr in LUX Xe.

The LUX 85.3~day WIMP search run background rate was elevated above expectations due to the presence of cosmogenically produced $^{127}$Xe. This isotope creates a low-energy ER background through the coincidence of low-energy X-ray generation and high-energy $\gamma$~ray de-excitation, where the $\gamma$~ray escapes detection by leaving the active region. This isotope decays with a 36~day half-life, and contributes an extra 0.5~mDRU$_{ee}$ to the 85.3~day WIMP search run backgrounds. The backgrounds generated by this isotope will not be present in future dark matter search runs.

Neutron emission rates from ($\alpha$,n) reactions, $^{238}$U fission, and high-energy muon interactions are predicted to create a subdominant NR background in LUX. A search was performed for low-energy MS events in the detector, as such events would be a signature of neutron scattering. No NR-like MS events below the 50\% NR acceptance mean were found during the 85.3~day run, consistent with predicted neutron emission rates. Neutron scatter rates within the WIMP search fiducial and energy regions are projected to be comparable between internal and external sources.

The ER S2/S1 band was characterized by high-statistics $^3$H calibration. The measured ER discrimination factor in LUX is 99.6\%, where NR events are characterized as falling below the NR S2/S1 band centroid.

Measured low-energy background rates are within 1$\sigma$ of expectation. An additional transient background during the first half of the WIMP search run was measured, in excess of expectations from $^{127}$Xe. The average background rate during the WIMP search run was $3.6\pm0.4$~mDRU$_{ee}$. 0.64~events are projected to fall below the NR centroid in the 85.3~day WIMP search data set, based on measured ER rates. One event was observed at the NR centroid, with none falling below. The data taken during the 85.3~day run show an overall agreement with the background-only model, with a p~value of 35\%.

The projected background rate for the 2014 one-year $\times$ 100~kg WIMP search run is $1.7\pm0.3$~mDRU$_{ee}$. The projected one-year run background rate is reduced by 55\% relative to 85.3~day rate due to the decay of all transient backgrounds, as well as the use of a smaller fiducial volume. Further reductions in background are expected in particular from optimization of the shape of the fiducial volume to minimize position-dependent background contributions. The model predicts a strong WIMP discovery potential for LUX for the upcoming one-year WIMP search run.

\bibliography{LUXBG_BibTeX}

%\begin{thebibliography}{99}

%% LUX NIM
%\bibitem{LUXNIM}D.~S. Akerib et al., NIM {\bf A704} (2013) 111-126.

%% LUX Run 3 PRL
%\bibitem{LUXPRL}D.~S. Akerib et al., \emph{submitted to PRL} (2013).

%% LUXSim
%\bibitem{LUXSim}D.~S. Akerib et al., NIM {\bf A675} (2012) 63-77.

%% NEST
%\bibitem{NEST}M. Szydagis et al., JINST {\bf 6} (2011) P10002.

%% R11410 screening
%\bibitem{R11410}D.~S. Akerib et al., NIM {\bf A703} (2013) 1-6.

%% Ti screening
%\bibitem{LUXTi}D.~S. Akerib et al., arXiv:1112.1376 (2012).

%% MENDL
%\bibitem{MENDL}MENDL reference!

%% ACTIVIA (the simulation, not the food)
%\bibitem{ACTIVIA}J.~J. Back and Y.~A. Ramachers, NIM {\bf A586} (2008) 286.

%% Copper activation at LNGS
%\bibitem{Laubenstein2009}M.~Laubenstein and G.~Heusser, \emph{Applied Radiation and Isotopes} {\bf 67} (2009) 750-754.

%% Copper activation review
%\bibitem{Cebrian2010}S.~Cebrian et al., Astropart. Phys. {\bf 33} (2010) 316-319.

%% Cosmic ray flux vs altitude
%\bibitem{Gordon2004}M.~S. Gordon et al., IEEE Transactions on Nuclear Science {\bf 51} (2004) 3427-3434.

%% LUX Run 2
%\bibitem{LUXSurfaceRun}D.~S. Akerib et al., Astropart. Phys. {\bf 45} (2013) 34-43.

%% LUX Kr removal paper
%\bibitem{LUXKrPaper}D.~S. Akerib e al., \emph{In preparation} (2013).

%% Kr detection in Xe
%\bibitem{KrDetInXe}A.~Dobi et al., NIM {\bf A665} (2011) 1-6.

%% Neutron Yield Tool
%\bibitem{NeutronYieldTool}D.M.~Mei and C.~Zhang, Neutron yield in materials, http://neutronyield.usd.edu/, 2008.

%% Neutron Yield Tool NIM
%\bibitem{NYTNIM}D.M.~Mei et al., NIM {\bf A606} (2009) 651.

%
%\end{thebibliography}

\end{document}